\documentclass[a4paper,11pt]{article}
\usepackage{jheppub}
\usepackage[most]{tcolorbox}
\usepackage{slashed}
\usepackage{amsmath}
\usepackage[most]{tcolorbox}
\usepackage{esint}
\usepackage{babel}
\usepackage{color}
\usepackage{enumitem}
\usepackage{subfigure}
\usepackage{float}
\usepackage{amsfonts}
\usepackage{amssymb}
\usepackage{amsthm}
\usepackage{enumitem}
\newcommand{\la}{\langle}
\newcommand{\ra}{\rangle}

\begin{document}

\title{Tree-Level Gravity Amplitudes at Infinity}

\author{Justin Lemmon,}
\author{Jaroslav Trnka}

\affiliation{Center for Quantum Mathematics and Physics (QMAP), University of California, Davis, 95616 CA, USA}

\emailAdd{jlemmon@ucdavis.edu}
\emailAdd{trnka@ucdavis.edu}

\abstract{In this note we study on-shell tree-level gravity amplitudes in the infinite momentum limit. In the case of the two-line BCFW shift, we have a famous improved behavior at infinity that allows for the amplitude to be reconstructed from the pole factorization. For other shifts, the poles at infinity are present and need to be considered, however general principles do not fix the residues of the amplitude on these poles. The web of all possible shifts is large, we focus primarily on a case of $(n{-}2)$-line anti-holomorphic shift, which also appears in the context of unitarity cuts of gravity loop integrands. We will find that for one class of shifts the gravity amplitudes at infinity exhibit a peculiar factorization property, quite different from the usual factorization on poles, while for other shifts, they evaluate to the same amplitude on shifted kinematics. We also discuss generalizations of our results to other anti-holomorphic shifts.}

\maketitle

\section{Introduction}

The study of graviton scattering amplitudes has been of great interest for many decades, both at the tree-level and at higher loops. This has led to many  discoveries including KLT relations \cite{Kawai:1985xq}, color-kinematics duality \cite{Bern:2008qj,Bern:2010ue,Bern:2019prr}, improved UV behavior in supergravity theories \cite{Bern:2008pv,Bern:2009kd,Bern:2018jmv,Bern:2023zkg}, recursion relations for tree-level amplitudes \cite{Britto:2004ap,Britto:2005fq,Cachazo:2005ca,Bedford:2005yy,Cohen:2010mi,Cheung:2015cba}, the CHY formalism \cite{Cachazo:2013gna,Cachazo:2013iea}, soft theorems \cite{He:2014laa,Cachazo:2014fwa,Campiglia:2014yka,Sen:2017nim}, on-shell diagrams and Grassmannians \cite{Herrmann:2016qea,Heslop:2016plj,Farrow:2017eol,Armstrong:2020ljm,Paranjape:2023qsq,Bourjaily:2023ycy}, and many others. There are also many fascinating closed form expressions for tree-level amplitudes in the literature, especially in the Maximal-Helicity-Violating (MHV) sector: starting from the Berends-Giele-Kuijf formula \cite{Berends:1988zp}, later proven to be equivalent to Mason-Skinner formula \cite{Mason:2009afn}; inverse soft factor representations \cite{Bern:1998sv,Nguyen:2009jk}, and the solution to the Britto--Cachazo--Feng--Witten (BCFW) recursion relations \cite{Drummond:2009ge}. Inspired by both the KLT formula and the BCFW recursion relations, a simple MHV expression was found in \cite{Elvang:2007sg}. Powerful code is also available that are able to generate explicit data to higher points for any helicity \cite{Bourjaily:2023uln}.

The most interesting expression is the Hodges formula \cite{Hodges:2012ym}, which expresses the amplitude as a certain determinant and exhibits the $S_n$ permutational symmetry of the $n$-point amplitude. The formula was later used in the formulation of the twistor string for ${\cal N}=8$ supergravity \cite{Cachazo:2012pz,Skinner:2013xp}, and can also be interpreted in the context of ambitwistor strings \cite{Geyer:2014fka}. Despite all these fascinating expressions, we know very little about any underlying geometric formulation of gravity amplitudes or any potential new symmetries --- which the Hodges formula strongly suggests. This is in contrast with our understanding of gluon scattering amplitudes, where the tree-level amplitudes enjoy dual conformal symmetry and Yangian invariance \cite{Drummond:2008vq,Drummond:2009fd}, the positive Grassmannian and on-shell diagrams \cite{ArkaniHamed:2009dn,ArkaniHamed:2009vw,Mason:2009qx,ArkaniHamed:2009sx,ArkaniHamed:2009dg,Arkani-Hamed:2016byb,Franco:2015rma,Paranjape:2022ymg}, and the Amplituhedron \cite{Arkani-Hamed:2013jha,Arkani-Hamed:2013kca,Arkani-Hamed:2017vfh,Arkani-Hamed:2021iya,Arkani-Hamed:2018rsk,Damgaard:2019ztj,Dian:2022tpf,Ferro:2022abq}. Some attempts in this direction have led to a new representation for NMHV gravity amplitudes \cite{Trnka:2020dxl} and established the uniqueness of MHV amplitudes from zeroes \cite{Koefler:2024pzv}, but no concrete geometric picture for graviton amplitudes is known to this day.

Apart from the explicit expressions, one of the most surprising features of gravity amplitudes is the improved $1/z^2$ scaling under the BCFW shift, which allows the computation of gravity amplitudes recursively \cite{Cachazo:2005ca,Bedford:2005yy,Hodges:2011wm}. While this scaling was proven from a detailed study of Feynman diagrams \cite{Arkani-Hamed:2008bsc}, it is fair to say that it is still not well understood. In contrast, the gluon amplitudes have a weaker $1/z$ fall-off at infinity which is linked to the dual conformal symmetry. No such interpretation (of even stronger large $z$ behavior!) is known in the case of graviton amplitudes. The absence of poles at infinity is very special for this particular shift, and does not generalize even to the next-to-simplest case: Risager three-line three-line shift \cite{Risager:2005vk}. In fact, the pole at infinity for this shift has been explicitly calculated for the 12-point NMHV amplitude, for which the Risager shift famously generates a single pole, in \cite{Conde:2012ik}. 
This was further generalized in \cite{Cachazo:2024mdn} to a general $n$-point NMHV amplitude and in \cite{Belayneh:2024lzq} for arbitrary N$^{k{-}2}$MHV amplitudes under a certain $k$-line anti-holomorphic shift. These results will also be relevant in this paper as certain boundary cases of our own analysis.

Note that the behavior of gravity amplitudes at large shifts critically depends on the number of legs we scale and the direction in which we send the momenta to infinity. This is in contrast to Yang-Mills amplitudes, where no non-trivial poles at infinity are ever present (apart from `bad shifts' due to helicity factors). The phase space of all such scalings, technically realized using momentum shifts, is huge, and we do not expect to get a universal answer for all of them. One particular choice is to do a multi-line shift, either holomorphic or anti-holomorphic. These shifts were classified for general QFTs in \cite{Cohen:2010mi}, where the Risager shift corresponds to the minimal case and the all-line shift is on the opposite side of the spectrum. In our paper, we study a general anti-holomorphic $m$-shift of the $n$-particle amplitude, mostly devoted to the case of $m=n{-}2$, but we do not restrict ourselves to shifting particular helicities. This is motivated by the unitarity cuts of four-point graviton loop amplitudes. On the multi-unitarity cut, the $L$-loop amplitude factorizes into a product of two tree-level amplitudes with two unshifted external momenta and $L{+}1$ shifted internal (on-shell) momenta. The scaling and behavior of these tree-level amplitudes at infinity directly feeds into the behavior of the cut integrand where loop momenta are sent to infinity (in a particular way), linking it to the associated UV structures in the $L$-loop amplitude. This particular shift and the correspondence with multi-unitarity cuts was first explored in \cite{Edison:2019ovj} based on earlier observations in \cite{Herrmann:2016qea,Herrmann:2018dja}. Surprisingly, the scaling of the cut is better than the contributions of individual Feynman diagrams in the loop amplitude, as a consequence of the special behavior of tree-level amplitudes under this multi-line shift.

\newpage

The organization of the paper is as follows: In Section 2, we briefly review some of the basic ingredients of tree-level graviton amplitudes. In Section 3, we motivate the discussion of poles at infinity by studying multi-loop unitarity cuts before introducing a multi-line shift and the limit to infinity. We also discuss the all-line and all-but-one line shifts, which happen to be very simple in our case. In Section 4, we analyze in details the $(n{-}2)$-line shift and discover two different patterns of the amplitude's behavior at infinity, which depend on the helicity configuration. In Section 5, we discuss various generalizations of our results to other shifts, and comparisons to results in the literature. We end with the conclusions and outlook. 

\section{Review of gravity amplitudes}

We now review basic ingredients and fundamentals we will need in the later discussions. 

\subsection*{Basics of tree-level amplitudes}

Tree-level graviton scattering amplitudes are gauge-invariant, on-shell rational functions of kinematical variables. In four dimensions, we can use \emph{spinor helicity variables} to encode the massless kinematics as
\begin{equation}
    p_k^\mu = \sigma^\mu_{a\dot{a}}\lambda_k^{(a)}\widetilde{\lambda}_k^{(\dot{a})} \, ,
\end{equation} 
where the Lorentz invariants are given by 
\begin{equation}
    \la ij\ra = \epsilon_{ab}\lambda_i^{(a)}\lambda_j^{(b)}\qquad   
    [ij] = \epsilon_{\dot{a}\dot{b}}\widetilde{\lambda}_i^{(\dot{a})}\widetilde{\lambda}_j^{(\dot{b})} \, .
\end{equation}
The spinor helicity variables are constrained to satisfy a momentum conservation, 
\begin{equation}
    \sum_k p_k^\mu = \quad\rightarrow \quad \sum_i \lambda_i^a\widetilde{\lambda}_i^{\dot{a}} = 0
\end{equation}
for $a,\dot{a}=1,2$, which gives four constraints, the same as the original version in the momentum space. We are interested in the $n$-point N$^{k{-}2}$MHV tree-level amplitude
\begin{equation}
{\cal M}_{n,k}(1^{h_1},2^{h_2},3^{h_3},{\dots},n^{h_n})(\lambda_1,\widetilde{\lambda}_1,\dots \lambda_n,\widetilde{\lambda}_n)
\end{equation}
where $h_i=\pm$ are helicities. The little group that leaves the momentum $p^\mu$ invariant acts as a simple rescaling of the spinor helicity variables: $\lambda_i\rightarrow t\lambda_i$, $\widetilde{\lambda}_i\rightarrow 1/t \,\widetilde{\lambda}_i$. Under this transformation, the positive and negative graviton helicities transform as
\begin{equation}
{\cal M}_n(\dots i^-\dots) \rightarrow t^4 {\cal M}_n(\dots i^-\dots),\quad {\cal M}_n(\dots j^+\dots) \rightarrow \frac{1}{t^4} {\cal M}_n(\dots j^+\dots)
\end{equation}
Note that the gravity amplitudes are unordered, so it does not matter in which order we write the indices of ${\cal M}_n$. The elementary amplitudes in the leading coupling (Einstein gravity) are three-point,
\begin{equation}
    {\cal M}_{3,1}(1^+,2^+,3^-) = \frac{[12]^6}{[13]^2[23]^2},\qquad 
    {\cal M}_{3,2}(1^-,2^-,3^+) = \frac{\la 12\ra^6}{\la 13\ra^2\la 23\ra^2} \, , 
\end{equation}
while the all-plus and all-minus three point amplitudes are non-zero only if we turn on the higher-derivative $F^3$ correction. At higher points, the form and complexity of the amplitude depend on graviton helicities. As usual, we use $k$ to denote the number of negative helicity gravitons, the number of positive helicity gravitons is $n{-}k$,
\begin{equation}
    {\cal M}_{n,k} (1^-,2^-,\dots,k^-,(k{+}1)^+,(k{+}2)^+,\dots, n^+)
\end{equation}
At four points there is only the MHV amplitude,
\begin{equation}
    {\cal M}_{4,2}(1^-,2^-,3^+,4^+) = \la 12\ra^8\cdot\frac{[34]}{\la12\ra^2\la13\ra\la14\ra\la 23\ra\la 24\ra\la 34\ra}, \label{MHV4}
\end{equation}
where we split the helicity factor $\la12\ra^8$ from the rest of the amplitude. Note that $[34]/\la12\ra$ is invariant under permutations of all labels, which then guarantees the permutational invariance of the amplitude (up to the helicity factor). The four-point amplitude is parity self-dual, and we can also write
\begin{equation}
     {\cal M}_{4,2}(1^-,2^-,3^+,4^+) = [34]^8\cdot \frac{\la12\ra}{[12][13][14][23][24][34]^2}.
\end{equation}
In this case, there is another formula which makes this self-duality manifest, ${\cal M}_4^{(2)}(1^-2^-3^+4^+) = \la12\ra^4[34]^4/stu$. At five points, we have MHV and NMHV amplitudes which are related by parity transformation $h^+\leftrightarrow h^-$, $\la ij\ra\leftrightarrow[ij]$. For the five-point MHV amplitude we have,
\begin{equation}
{\cal M}_{5,2}(1^-,2^-,3^+,4^+,5^+) = \frac{\la12\ra^8\cdot {\rm Tr}(1234)}{\la12\ra\la13\ra\la14\ra\la15\ra\la23\ra\la24\ra\la25\ra\la34\ra\la35\ra\la45\ra} \label{MHV5}
\end{equation}
where the trace is defined as
\begin{equation}
{\rm Tr}(1234) \equiv {\rm Tr}(\slashed{p_1}\slashed{p_2}\slashed{p_3}\slashed{p_4})=  \la12\ra[23]\la34\ra[41]-[12]\la23\ra[34]\la41\ra
\end{equation}
The trace is completely antisymmetric in all five labels, so ${\rm Tr}(1234)=-{\rm Tr}(1235)$, etc. At higher points, there are multiple independent amplitudes for each $n$. It is worth emphasizing that the permutational symmetry of the amplitude is a very strong constraint and this property can not be made manifest in the spinor helicity variables as evident from (\ref{MHV4}), (\ref{MHV5}). Different permutations of the same formula can be shown to be equivalent only when the momentum conservation is used. This is in contrast with the color-ordered tree-level Yang-Mills amplitudes where we can use dual variables (or momentum twistors) which make the cyclic symmetry manifest. The absence of the permutationally invariant kinematical variables is one of the main reasons the theoretical understanding of gravity amplitudes is well behind that of their gluon counterparts. We refer readers to excellent books and reviews on this topic \cite{Elvang:2013cua,Dixon:2013uaa,Cheung:2017pzi,Travaglini:2022uwo,Badger:2023eqz}.

There are several fascinating explicit formulas for all-multiplicity graviton amplitudes in the literature. The most remarkable expression is the Hodges formula \cite{Hodges:2012ym} for MHV amplitudes,
\begin{equation}
    {\cal M}_{n,2}(1^-,2^-,3^+,{\dots},n^+) = \la 12\ra^8\times\frac{{\rm det}(\Phi_{abc}^{def})}{(abc)(def)} \label{Hodges1} \, ,
\end{equation}
where we denoted $(abc)=\la ab\ra\la bc\ra\la ca\ra$ and $\Phi_{abc}^{def}$ is the $(n{-}3\,\times\,n{-}3)$ matrix which is obtained from the $n\times n$ matrix $\Phi$ by deleting rows $a,b,c$ and columns $d,e,f$. The elements of the $\Phi$ matrix are defined as
\begin{equation}
    \Phi_{i,j} = \frac{[ij]}{\la ij\ra},\qquad \Phi_{i,i} = - \sum_{j\neq i}\frac{[ij]\la ja\ra\la jb\ra}{\la ij\ra \la ia\ra\la ib\ra} \label{Hodges2} \, ,
\end{equation}
where $\lambda_a$, $\lambda_b$ are arbitrary spinors. This formula makes the $S_n$ permutational symmetry manifest, but any choice of $a,b,c,d,e,f$ would break it, so there is no permutationally symmetric expression in the spinor helicity space. We can use this formula to write a compact expression for the six-point amplitude:
\begin{align}
    &{\cal M}_{6,2}(1^-,2^-,3^+,4^+,5^+,6^+) = \frac{\la 12\ra^8}{\la12\ra\la23\ra\la13\ra\la45\ra\la46\ra\la56\ra}\\
    &\hspace{-0.2cm}\times \Bigg\{\frac{[14]}{\la14\ra}\left[\frac{[25]}{\la25\ra}\frac{[36]}{\la36\ra} {-} \frac{[26]}{\la26\ra}\frac{[35]}{\la35\ra}\right]-\frac{[15]}{\la15\ra}\left[\frac{[24]}{\la24\ra}\frac{[36]}{\la36\ra} {-} \frac{[26]}{\la26\ra}\frac{[34]}{\la34\ra}\right]-\frac{[16]}{\la16\ra}\left[\frac{[25]}{\la25\ra}\frac{[34]}{\la34\ra} {-} \frac{[24]}{\la24\ra}\frac{[35]}{\la35\ra}\right]\Bigg\}\nonumber  
\end{align}
There are also other interesting all-multiplicity expressions for MHV amplitudes \cite{Berends:1988zp,Mason:2009afn,Elvang:2007sg,Bern:1998sv,Nguyen:2009jk} and also some suggestive formulas for NMHV amplitudes based on on-shell diagrams \cite{Bourjaily:2023uln,Paranjape:2023qsq}.

\subsection*{On-shell recursion relations}

The on-shell recursion relations are based on the simple idea of reconstructing the $n$-point amplitude from factorization channels. Based on tree-level unitarity, any tree-level amplitude ${\cal M}_n$ factorizes on the pole $P^2=0$, where $P=\sum_{i\in \sigma} p_i$ is the sum of a subset of momenta, into the product of two tree-level amplitude ${\cal M}_L$ and ${\cal M}_R$, 
\begin{equation}
	\begin{tabular}{cc}
	 \includegraphics[scale=.7]{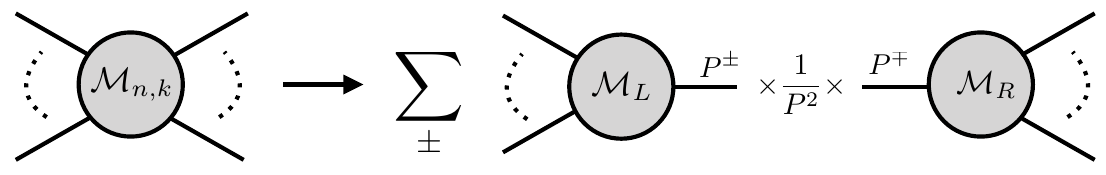} \label{fact}
	\end{tabular} 
\end{equation}
where we have to sum over the helicities of the internal on-shell leg with momentum $P$. If the collection of all such conditions is enough to fix the amplitude uniquely --- if it is a unique kinematical function which factorizes properly on all poles --- we can reconstruct it using factorization channels. This idea is technically implemented by \emph{recursion relations}. In this setup, we perform a momentum shift on some (or all) external kinematical variables. For spinor helicity variables, 
\begin{equation}
    \lambda_i \rightarrow \lambda_i(z),\quad \widetilde{\lambda}_i\rightarrow \widetilde{\lambda}_i(z)
\end{equation}
provided the shifted variables still satisfy momentum conservation. As a result, we get a shifted amplitude ${\cal M}_n(z)$ which depends on the original spinor helicity variables $\lambda_i$, $\widetilde{\lambda}_i$ and the shift parameter $z$. We can also use multiple shift parameters (and we will do so in the case of our interest), but we stick to one parameter here for simplicity of the explanation.

The simplest case is the two-line shift on the legs $i$ and $j$. The form is completely fixed by the momentum conservation,
\begin{equation}
    \lambda_i(z) = \lambda_i + z\lambda_j,\quad \widetilde{\lambda}_i(z)=\widetilde{\lambda}_i - z \widetilde{\lambda}_j \label{BCFW}
\end{equation}
which is the Britto-Cachazo-Feng-Witten (BCFW) shift. The idea now is to study the shifted amplitude ${\cal M}_n(z)$ as a function of only the parameter $z$ and leave the kinematical variables frozen. Because all poles of ${\cal M}_n(z)$ in $z$ come from the standard poles $1/P^2$ of the original unshifted amplitude, we can access these (now shifted) poles $1/P^2(z)$ by fixing the parameter $z$ to a special value $z^\ast$ such that 
\begin{equation}
    P^2(z^\ast)=0
\end{equation}
while keeping the spinor helicity generic. Then the shifted amplitude factorizes on the pole,
\begin{equation}
    {\cal M}_n(z) \xrightarrow{z=z^\ast} \frac{z^\ast}{z-z^\ast}\frac{{\cal M}_L(z^\ast){\cal M}_R(z^\ast)}{P^2}
\end{equation}
where both subamplitudes ${\cal M}_{L,R}$ are now evaluated at \emph{deformed kinematics} $\lambda_i(z^\ast)$, $\widetilde{\lambda}_i(z^\ast)$. The shifted amplitude ${\cal M}_n(z)$ has multiple poles --- in the generic case as many as factorization channels of the unshifted amplitude ${\cal M}_n$ -- and has a schematic form,
\begin{equation}
{\cal M}_n(z) = \frac{n(z)}{\prod_k (z-z^\ast_k)}
\end{equation}
where the numerator $n(z)$ is a polynomial in $z$ of the degree $d$. We consider the following contour integral in the complex $z$-plane 
\begin{equation}
    \oint \frac{dz}{z} {\cal M}_n(z) = 0 \label{GRT}
\end{equation}
where the contour is taken to encircle all poles of ${\cal M}_n(z)$ and the origin $z=0$. The pole at origin is added for purpose because the original unshifted amplitude is 
\begin{equation}
  {\cal M}_n = \underset{z=0}{\rm Res}\,\frac{{\cal M}_n(z)}{z}
\end{equation}
denoting $P_k=\sum_{i\in k} p_i$ and using (\ref{GRT})
\begin{equation}
    {\cal M}_n = - \sum_k \underset{z=z_k^\ast}{\rm Res}\,\frac{{\cal M}_n(z)}{z} = - \sum_k \frac{{\cal M}_L(z_k^\ast){\cal M}_R(z_k^\ast)}{P_k^2}
\end{equation}
where we sum over all poles of the shifted amplitude.  This is the standard form of the \emph{BCFW recursion relations}, which reproduce the $n$-point amplitude ${\cal M}_n$ in terms shifted lower-point amplitude ${\cal M}_L$, ${\cal M}_R$. We can recurse this formula all the way down to the elementary three-point amplitudes.

Note that the poles of the shifted amplitude are just a subset of all factorization channels. Some of the poles of ${\cal M}_n$ stay unshifted after performing the BCFW shift (these are the poles where both legs $i,j$ are on the same side of the factorization channel), and hence do not appear in the recursion relations. This residue theorem only works if the shifted amplitude has \emph{no poles at infinity}, in other words if
\begin{equation}
    {\cal M}_n(z) \sim \frac{1}{z} \quad\mbox{or better for}\,\,z\rightarrow\infty
\end{equation}
This is true for Yang--Mills amplitudes which scale as $1/z$, but very surprisingly this is also true for gravity amplitudes that scale even better as $1/z^2$. While this improved behavior was understood at the level of Feynman diagrams \cite{Arkani-Hamed:2008bsc}, a deeper understanding is still missing. For Yang--Mills amplitude the absence of a pole at infinity is closely related to the dual conformal symmetry of the tree-level gluon amplitudes, but no such symmetry explanation for the gravity scaling --- which is even better --- is known.

Gravity amplitudes do have poles at infinity, these poles are simply avoided when doing the BCFW shift. When performing other shifts we can easily access them. The next-to-simplest shift is the three-line shift. If we disregard a possibility of two successive BCFW shifts, the three-line shift is unique (up to parity transformation) and takes the form
\begin{equation}
\widetilde{\lambda}_1(z)=\widetilde{\lambda}_1+z\la23\ra\widetilde{\eta},\qquad
\widetilde{\lambda}_2(z)=\widetilde{\lambda}_2+z\la31\ra\widetilde{\eta},\qquad
\widetilde{\lambda}_3(z)=\widetilde{\lambda}_3+z\la12\ra\widetilde{\eta}. \label{Risager}
\end{equation}
where $\widetilde{\eta}$ is an arbitrary spinor. This is known as the \emph{Risager shift} \cite{Risager:2005vk}. The recursive procedure is similar to that of the BCFW shift, though more terms appear in the recursion as more poles $1/P^2$ of the original amplitude are shifted. Furthermore, the individual terms in the recursion depend on the arbitrary spinor $\widetilde{\eta}$ while the final result is independent of it. However, the most important difference is the scaling of the shifted amplitude ${\cal M}_n(z)$ at infinity. The MHV amplitudes do not generate any pole (as in the BCFW shift) but the NMHV amplitude scales like
\begin{equation}
    {\cal M}_{n,3}(z) \sim \frac{1}{z^{12-n}}
\end{equation}
and for $n=12$ this scaling produces a pole in the contour integral (\ref{GRT}) which was first observed in \cite{Bianchi:2008pu}. This creates an interesting problem because unlike the $1/P^2$ poles, the behavior of the amplitude on the pole at infinity is not predicted by the tree-level unitarity and we do not have a formula like (\ref{fact}). So unless we find a way how to calculate these poles, the recursive formula can not be used. On the other hand, one can take a position that these poles are all irrelevant because we do have the BCFW recursion relations which avoids them. While this is true for a practical reason of the computation of tree-level amplitudes, the presence of such poles provides us with an opportunity to understand them, calculate and further used this knowledge in different situations. 

It is also worth noting that in a very different setup of amplitudes, in effective field theories, poles at infinity exist as well and here even the standard BCFW recursion relations does not work. However, for a special class of theories, called exceptional EFTs \cite{Cheung:2014dqa,Cheung:2016drk,Cheung:2018oki,Elvang:2018dco}, the amplitude exhibits a very special behavior in the soft limit. One can then exploit this fact and modify the contour integral (\ref{GRT}) (by adding more $z$-dependent terms in the denominator) such that even the dangerous scaling of ${\cal M}_n(z)$ at infinity does not produce new poles. This procedure is called soft recursion relations \cite{Cheung:2015ota}.

\section{Multi-line shifts}

The BCFW and Risager shifts are the unique 2-line and 3-line shifts with very different behavior at infinity in the case of gravity amplitudes. Our aim will be to study multi-line shifts, but there is a very broad phase space of all possibilities one can consider. We will focus on the multi-line anti-holomorphic shift, motivated by the study of cuts of loop integrands.

\subsection{Motivation: multi-unitarity cuts}

As we have eluded before, we take our inspiration to study poles at infinity of tree-level gravity from loop amplitudes. In particular, we are inspired by the analysis of poles at infinity of the $L$-loop amplitude where we send the loop momenta $\ell_i\rightarrow\infty$ for $i=1,2,{\dots},L$. This is the region where UV divergencies appear when integrating over all loop momenta. 

The $L$-loop amplitude is given by the sum of Feynman diagrams or scalar integrals multiplied by gauge invariant coefficients (in the context of unitarity methods), integrated over $L$ off-shell loop momenta,
\begin{equation}
    {\cal M}^{(L)}_n = \sum_k I_k \quad \mbox{where} \quad I_k = \int d^4\ell_1\,d^4\ell_2\dots d^4\ell_L\,{\cal I}_k \label{integral}
\end{equation}
In planar theories, we can choose global loop momenta for all contributing terms, switch the sum and integral and write 
\begin{equation}
    {\cal M}^{(L)}_n = \int d^4\ell_1\,d^4\ell_2\dots d^4\ell_L\,{\cal I}_n^{(L)}
\end{equation}
where we integrate over the Minkowski contour, ie. all loop momenta are real. The object ${\cal I}_n^{(L)}$ is a \emph{planar integrand}. In this case, it makes sense now to approach the limit where all loop momenta $\ell_i\rightarrow\infty$ and study the pole. The absence of such pole in the ${\cal N}=4$ SYM theory is closely related to the presence of the dual conformal symmetry, and it is a seed for further mathematical constructions \cite{Arkani-Hamed:2010zjl}. But in other planar gauge theories, these poles are present and are definitely interesting to study! Unfortunately, this is not possible in non-planar theories (such as gravity) where the global loop variables do not exist (though interesting progress on this question has been recently made in \cite{Arkani-Hamed:2023lbd,Arkani-Hamed:2023mvg,Arkani-Hamed:2023jry,Bern:2024vqs}). There we can only talk about poles at infinity of individual Feynman diagrams, but not the full amplitude, because the loop momenta are not lined up between different diagrams. 

However, instead of the loop amplitude ${\cal M}^{(L)}_n$ we can study a different object: \emph{cut of the loop amplitude}, which can be thought of as a change of the integration contour where some of the degrees of freedom in the loop momenta $\ell_1,{\dots},\ell_L$ are integrated over $S^1$ contours encircling poles, while the remaining ones are still integrated over the Minkowski contour. 
\begin{equation}
 \int d^4\ell_1{\dots} d^4\ell_L \rightarrow 
 \underbrace{ \oint_{S^1}\dots \oint_{S^1}}_{p\,\text{times}} \int d^{4L{-}p}\ell
\end{equation}
The integration over $S^1$ contours reduces to residues on poles of (\ref{integral}) which are propagators $1/P^2$ where $P$ is now a function of both external and loop momenta. This is a well-defined procedure even in the absence of a single integrand function, as long as all loop momenta are cut. Unitarity then dictates that the result is a product of tree-level amplitudes. The minimal number of propagators to cut is $p=(L{+}1)$ at $L$-loops and the corresponding object is the \emph{multi-unitarity cut},
\begin{equation}
	\begin{tabular}{cc}
	 \includegraphics[scale=.67]{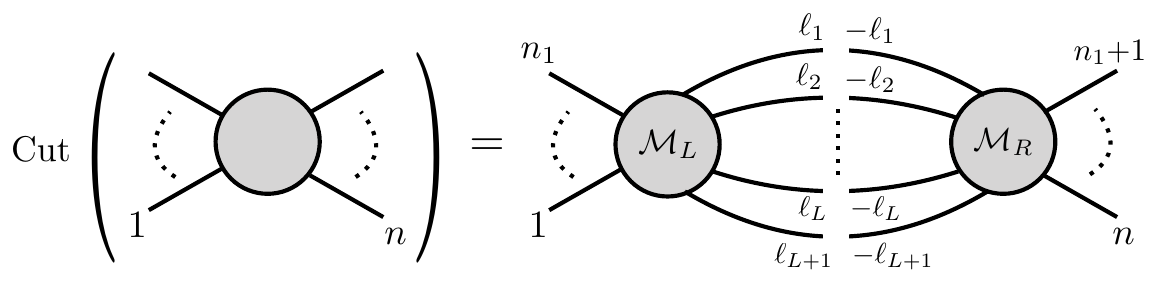}
	\end{tabular} 
\end{equation}
In this case, the loop amplitude factorizes into the product of two tree-level amplitudes,
\begin{equation}
    {\rm Cut}\,{\cal M}^{(L)}_n = \sum_\pm {\cal M}_L^{(0)}(1,2,{\dots},n_1,\ell_1^\pm,{\dots},\ell_{L{+}1}^\pm) \times {\cal M}_L^{(0)}(-\ell_1^\mp,{\dots},-\ell_{L{+}1}^\mp,n_1{+}1,{\dots},n) \label{multi}
\end{equation}
where the momentum conservation solves for the momentum $\ell_{L{+}1}$ in terms of others,
\begin{equation}
    p_1+p_2+\dots+p_{n_1}+\ell_1+\ell_2+\dots+\ell_L+\ell_{L{+}1}=0
\end{equation}
The sum is over the helicities of the internal particles $\ell_1,{\dots},\ell_{L{+}1}$ (which are opposite for $\ell_k$ and $-\ell_k$ legs). This is the cut we will be inspired by when choosing particular shifts of tree-level amplitudes to study.

It is worth emphasizing that this cut localizes $L{+}1$ degrees of freedom on the non-Minkowski cut contour while remaining $3L{-}1$ degrees of freedom are still integrated over the Minkowski space. The cut information allows us to learn about certain aspects of the fully integrated amplitude, but the connection is not straightforward as the contours are different. The maximal $4L$-dimensional residues of the amplitudes (all degrees of freedom of loop momenta are localized on cuts), called \emph{leading singularities} \cite{Cachazo:2008vp}, appear as coefficients of transcendental functions and in planar ${\cal N}=4$ SYM theory are invariant under the Yangian symmetry. The multi-unitarity cuts are on the opposite side of the spectrum with the minimum number of propagators on-shell. Multi-unitarity cuts then preserve most of the information of the original (off-shell) loop amplitude. 

The multi-unitarity cuts in ${\cal N}=8$ SUGRA were studied extensively in \cite{Edison:2019ovj}, including the scaling at infinity. This was then compared to the contribution of individual Feynman diagrams (or scalar integrals in the context of unitarity methods). Very surprisingly scaling of individual terms was worse than the behavior of the cut, which required massive cancelations between Feynman diagrams in the full sum. This gave an evidence of potential surprising UV behavior of gravity amplitudes (though this study was at the level of integrands and cuts, so the connection is indirect). From a different perspective, the poles at infinity were also studied in the context of other cuts of amplitudes, even in the ${\cal N}<4$ SYM theory \cite{Bourjaily:2021ujs,Brown:2022wqr}

To conclude this motivational section, we will reiterate that the multi-unitarity cut is a very interesting on-shell gauge invariant object, defined as a sum of products of tree-level amplitudes (\ref{multi}). It provides non-trivial information about multi-loop amplitudes, and the poles at infinity encode some data about the UV sector of the theory. The kinematical properties of the multi-unitarity cut can be in principle traced back to individual tree-level amplitudes in (\ref{multi}). To study the behavior of a single tree-level amplitude at infinity is the aim of this paper. 

\subsection{Multi-line shifts and approaching infinity}

The central object of our interest is the $n$-point tree-level graviton amplitude, where we distinguish $n_1$ particles with unshifted momenta and standard labels $1,2,\dots,n_1$, and $n_2$ particles with shifted momenta and hatted labels $n_1{+}1,\dots,n$, where $n=n_1{+}n_2$.  
\begin{equation}
{\cal M}_{n,k}(1,2,\dots,n_1,\widehat{n_1{+}1},\widehat{n_1{+}2},\dots,\widehat{n})
\end{equation}
Let us denote now $k_1$ the number of negative helicity gravitons among the unshited legs and $k_2$ the number of negative helicity gravitons of  shifted legs, with $k_1+k_2=k$ of the $n$-point N$^{k{-}2}$MHV amplitude ${\cal M}_{n,k}$. We use a more convenient notation that makes the bookkeeping simpler (rather than listing helicities of all legs and which are shifted/unshifted) in the following, 
\begin{align}
&{\cal M}_{n,k}(\{k_1,n_1{-}k_1\}) \equiv \\
&\hspace{1.6cm} {\cal M}_{n,k}(1^-,2^-,\dots,k_1^-,(k_1{+}1)^+,\dots,n_1^+,\widehat{n_1{+}1}^-,\dots,\widehat{n_1{+}k_2}^-,\widehat{n_1{+}k_2{+}1}^+,\dots,\widehat{n}^+) \nonumber \, ,
\end{align}
where $k_1$ is the number of unshifted negative helicity gravitons and $n_1-k_1$ is the number of unshifted positive helicity gravitons.
\begin{equation}
	\begin{tabular}{cc}
	 \includegraphics[scale=.7]{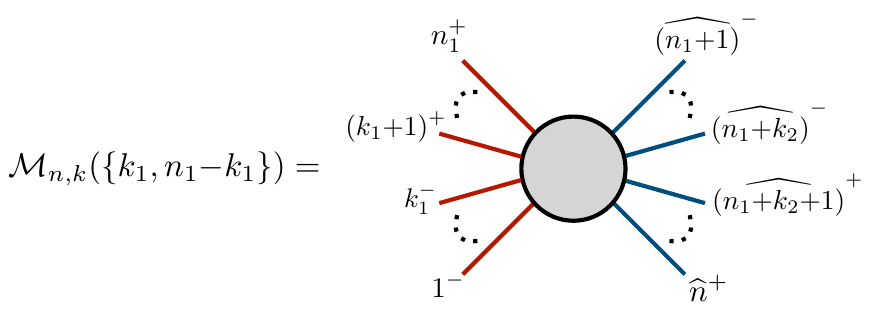}
	\end{tabular} 
\end{equation}
We perform a multi-line anti-holomorphic shift on the hatted momenta,
\begin{equation}
\widetilde{\lambda}_{\widehat{a}}(z) =  \widetilde{\lambda}_a + z\, \alpha_a\widetilde{\eta} \quad \mbox{for}\,\,a=n_1{+}1,\dots,n \label{shift}
\end{equation}
where $z$ and $\alpha_a$ are coefficients, $\widetilde{\eta}$ is an auxiliary spinor. The momentum conservation puts constraints on $\alpha_k$,
\begin{equation}
    \sum_{a=n_1{+}1}^n \alpha_a \lambda_a = 0 \label{momcons}
\end{equation}
This is an anti-holomorphic shift, so all the angle brackets are unchanged while the square brackets between two shifted legs become
\begin{equation}
[\hat{a}\hat{b}] = [(\widetilde{\lambda}_a{+}z\alpha_a\widetilde{\eta})(\widetilde{\lambda}_b{+}z\alpha_b\widetilde{\eta})] = [ab] +  z(\alpha_b[a\tilde{\eta}] {-} \alpha_a[b\tilde{\eta}])  \label{brackets1}
\end{equation}
For the brackets between one shifted leg $\hat{a}$ and one unshifted leg $b$ are now
\begin{equation}
    [a\hat{b}] = [ab] + z\alpha_b[a\widetilde{\eta}] \label{brackets2}
\end{equation}
Obviously for two unshifted legs the brackets $[ab]$ are unchanged. Now, we can approach infinity by performing the shift (\ref{shift}) and taking $z\rightarrow\infty$. We can now expand
\begin{equation}
{\cal M}_{n,k}(z) = z^m\,{\cal M}_{\rm inf} + {\cal O}(z^{m{-}1})\quad \mbox{in the limit}\,\,\, z\rightarrow \infty
\end{equation}
for some integer $m$. The leading behavior term at infinity ---  the amplitude at infinity ${\cal M}_{\rm inf}$ --- is a function we want to determine. It obviously depends on helicites of particles, both shifted and unshifted. This is a very large phase space of all possible values of $n_1,n_2,k_1,k_2$. Our bigger goal is to find a general behavior for any shift and understand the behavior at infinity on general grounds as a consequence of unitarity or other basic properties, similar to tree-level unitarity and factorizations for $P^2=0$ poles. 

\subsection{All line and all-but-one line shifts}

We can easily analyze two special cases: all line or $n$-line shift, and the all-but-one or $(n{-}1)$-line shift. For the $n$-line shift all legs are shifted,
\begin{equation}
    {\cal M}_{n,k}(\{0,0\}) = {\cal M}_{n,k}(\widehat{1}^-,\widehat{2}^-,\dots,\widehat{k}^-,\widehat{(k{+}1)}^+,\dots,\widehat{n}^+) \label{amp1}
\end{equation}
All square brackets are shifted and for $z\rightarrow\infty$ we see from (\ref{brackets1}),
\begin{equation}
[\hat{a}\hat{b}] = z(\alpha_b[a\tilde{\eta}] {-} \alpha_a[b\tilde{\eta}])  \label{brackets3}
\end{equation}
From that we can define new prime labels $a'$ with spinor helicity variables,
\begin{equation}
    \lambda_{a'}\equiv \lambda_a,\quad \widetilde{\lambda}_{a'} \equiv \left(\begin{matrix}\alpha_a \\ [a\tilde{\eta}] \end{matrix}\right)\label{prime1}
\end{equation}
It is easy to see that these spinors do satisfy momentum conservation,
\begin{equation}
    \sum_{a=1}^n \lambda_{a'}\widetilde{\lambda}_{a'} = \left(\begin{matrix}\sum_a\alpha_a \lambda_a \\ \sum_a\lambda_a[a\tilde{\eta}] \end{matrix}\right) = 0
\end{equation}
The angle brackets stay unchanged, $\la a'b'\ra=\la ab\ra$, while the square brackets are now
\begin{equation}
  [\hat{a}\hat{b}] = z [a'b'] \label{prime2}
\end{equation}
Importantly, the $n$-point N$^{k{-}2}$MHV amplitude ${\cal M}_{n,k}$ is homogeneous in the angle and square brackets 
\begin{equation}
    {\cal M}_{n,k} \sim \la\cdots\ra^{2k-n+1}[\cdots]^{n-2k+1}
\end{equation}
which follows from the mass dimension and little group scaling of ${\cal M}_{n,k}$. For the $n$-line shift all angle brackets are shifted and hence all of them scale as $z$ and produce (\ref{prime2}). As a result, the leading term at infinity is 
\begin{equation}
{\cal M}_{n,k}(\{0,0\} )= z^{n-2k+1} \times {\cal M}_{\rm inf} + {\cal O}(z^{n-2k}) \label{inf1}
\end{equation}
where the scaling was also pointed out in \cite{Cohen:2010mi}, and
\begin{equation}
    {\cal M}_{\rm inf} = {\cal M}_{n,k}(1',2',\dots,n') \label{inf2}
\end{equation}
The amplitude at infinity of ${\cal M}_{n,k}$ is the same amplitude, just with new spinor helicity variables (\ref{prime1}). A very similar argument applies to the $(n{-}1)$ line shift. The amplitude for negative unshifted graviton is
\begin{equation}
    {\cal M}_{n,k}(\{1,0\}) = {\cal M}_{n,k}(1^-,\widehat{2}^-,\dots,\widehat{k}^-,\widehat{(k{+}1)}^+,\dots,\widehat{n}^+) \label{amp2}
\end{equation}
while the for positive unshifted graviton is
\begin{equation}
    {\cal M}_{n,k}(\{0,1\}) = {\cal M}_{n,k}(1^+,\widehat{2}^-,\dots,\widehat{(k{+}1)}^-,\widehat{(k{+}2)}^+,\dots,\widehat{n}^+)\label{amp3}
\end{equation}
In both cases, two shifted legs will give the same brackets (\ref{brackets3}) as before, and hence the $\widetilde{\lambda}$ spinors are again given by (\ref{prime1}). For the unshifted label 1, the angle brackets obviously stay the same and the square brackets are now
\begin{equation}
    [1\hat{a}] = [1a] + z\alpha_a[1\widetilde{\eta}] \xrightarrow{z\rightarrow\infty} z \alpha_a[1\widetilde{\eta}] \label{brackets4}
\end{equation}
Hence we can define a new spinor $\widetilde{\lambda}_{1'}$ which is given by
\begin{equation}
   \lambda_{1'}\equiv \lambda_1,\quad \widetilde{\lambda}_{1'} \equiv \left(\begin{matrix} 0 \\ [1\tilde{\eta}] \end{matrix}\right)\label{prime3}
\end{equation}
for which we get at $z\rightarrow\infty$
\begin{equation}
    [1\hat{a}] = z [1'a']
\end{equation}
We can check that the momentum conservation is indeed satisfied for (\ref{prime1}), (\ref{prime3}). As a result, all square brackets $[\dots]$ for all labels scale as ${\sim}z$, and for both cases (\ref{amp2}), (\ref{amp3}) the amplitude at infinity is again the same amplitude just with shifted kinematics (\ref{prime1}), (\ref{prime3}) and we get the same expressions (\ref{inf1}), (\ref{inf2}) also for the $(n{-}1)$ line case.

If we have fewer shifted legs, the same argument does not apply, because we get brackets among unshifted spinors $[ab]$ which are unchanged for $z\rightarrow\infty$ and hence just scale as ${\sim}z^0$. Then the large $z$ limit disfavors these brackets in the competition with shifted brackets which scale as ${\sim}z^1$. Hence, we do not expect the scaling to be simply $z^{n-2k+1}$ and the amplitude at infinity ${\cal M}_{\rm inf}$ to be equal to the original amplitude ${\cal M}_{n,k}$ at shifted kinematics. More detailed analysis is needed on a case-by-case basis.

\section{Amplitudes at infinity for the $(n{-}2)$ shift}

In this section, we will focus on the first non-trivial multi-line shift when we shift $(n{-}2)$ external legs and we leave two legs unshifted. In our notation this is $n_1=2$. This is also a primary case of our interest from the point of view of the multi-unitarity cut as these amplitudes are relevant for the multi-loop $2\rightarrow2$ scattering amplitude,
\begin{equation}
	\begin{tabular}{cc}
	 \includegraphics[scale=.65]{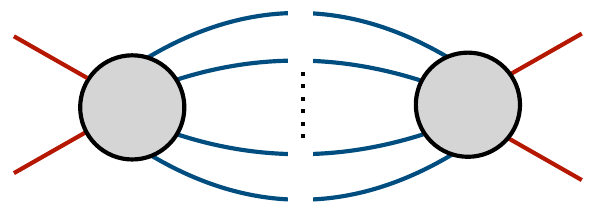}
	\end{tabular} 
\end{equation}
For fixed external legs, amplitudes of all helicity degrees contribute to each of the amplitude on the cut as the internal lines can have arbitrary helicities and we have to sum over them on the cut. As we will see, the result will mainly depend on the of helicities of the unshifted legs. The possible configurations are:
\begin{equation}
	\begin{tabular}{cc}
	 \includegraphics[scale=.72]{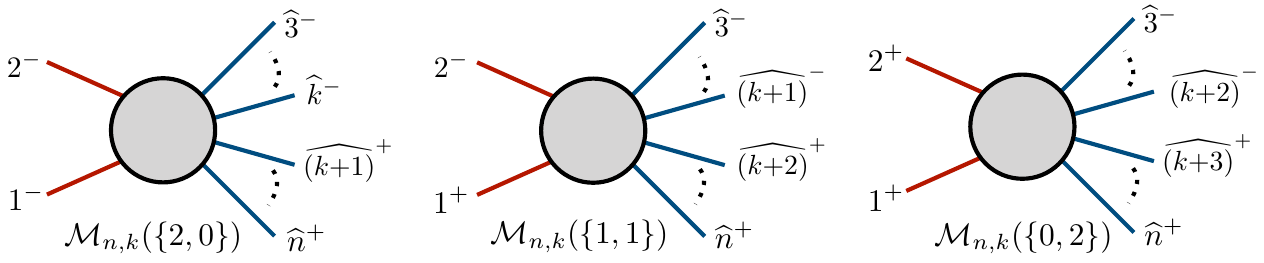} \label{exthel}
	\end{tabular} 
\end{equation}
All legs but two are shifted, and in $z\rightarrow\infty$ limit all square brackets scale like ${\sim}z$ from (\ref{brackets3}), (\ref{brackets4}) except the bracket $[12]$ which is unshifted and behaves like ${\sim}\,z^0$. This makes the analysis more complicated, as eluded to before, and we will do it on the case-by-case basis. 

\subsection{$\rm MHV$ amplitudes}

In the case of MHV amplitudes we use an advantage of the Hodges formula (\ref{Hodges1}) and choose $(a,b,c)=(d,e,f)=(1,2,3)$. We are agnostic about the distribution of negative helicities, whether they are among the shifted labels $\hat{1}$, $\hat{2}$ or not -- this is because of the trivial helicity factor that does not depend on the shift. As a result, we get for the shifted amplitude,
\begin{equation}
    {\cal M}_{n,2}(1,2,\widehat{3},\widehat{4},\dots,\widehat{n}) = \frac{\la ij\ra^8}{\la12\ra^2\la23\ra^2\la13\ra^2} \times {\rm det}(\Phi_{123}^{123}) \label{Hodges3}
\end{equation}
where now all the non-diagonal elements of $\Phi_{a,b}$ only depend on shifted labels $4,5,\dots,n$ and are given by
\begin{equation}
    \Phi_{a,b} = \frac{[\hat{a}\hat{b}]}{\la ab\ra} \rightarrow z\frac{[a'b']}{\la a'b'\ra}
\end{equation}
where $a',b'$ are given by the same prime spinors (\ref{prime1}). The diagonal element now combines both the shifted and unshifted square brackets
\begin{equation}
    \Phi_{a,a} = -\frac{[\hat{a}1]\la 1x\ra\la 1y\ra}{\la i1\ra\la ix\ra\la iy\ra}-\frac{[\hat{a}2]\la 2x\ra\la 2y\ra}{\la i2\ra\la ix\ra\la iy\ra} - \sum_{b=4}^n \frac{[\hat{a}\hat{b}]\la jx\ra\la jy\ra}{\la ab\ra \la ix\ra\la iy\ra} 
\end{equation}
where $x,y$ are arbitrary fixed spinors and in the sum the term $a=b$ is excluded. Note that the label $a$ only runs over $a=4,\dots,n$ which was the point of the special choice of (\ref{prime1}). For $z\rightarrow\infty$ we get for the square brackets
\begin{equation}
    [\hat{a}1]\rightarrow z\alpha_a [a1],\quad  [\hat{a}2]\rightarrow z\alpha_a [a2],\quad [\hat{a}\hat{b}]\rightarrow z(\alpha_a[b\widetilde{\eta}]-\alpha_b[a\widetilde{\eta}])
\end{equation}
so all scale like ${\sim}\, z$ and the diagonal element $\Phi_{a,a}$ keeps the current form 
\begin{equation}
    \widehat{\Phi}_{a,a} \rightarrow z\,\Phi_{a,a}(1',2',3',{\dots},n')
\end{equation}
where now the $\widetilde{\lambda}$ spinors for the prime labels are
\begin{equation}
     \widetilde{\lambda}_{1'} \equiv \left(\begin{matrix}0 \\ [1\tilde{\eta}] \end{matrix}\right),\quad   \widetilde{\lambda}_{2'} \equiv \left(\begin{matrix}0 \\ [2\tilde{\eta}] \end{matrix}\right),\quad
     \widetilde{\lambda}_{a'} \equiv \left(\begin{matrix}\alpha_a \\ [a\tilde{\eta}] \end{matrix}\right)\label{prime4} \, ,
\end{equation}
where $a=3,\dots,n$. Note that this has the same form as the all-line shift prime spinors (\ref{prime1}) with $\alpha_1=\alpha_2=0$. The $\lambda$ spinors stay unchanged and $\lambda_{a'}=\lambda_a$ for all $a=1,\dots,n$. Momentum conservation is again satisfied for these spinors, and the leading term at infinity is
\begin{equation}
{\cal M}_{n,2} = z^{n-3} \times {\cal M}_{\rm inf} + {\cal O}(z^{n-4})\quad\mbox{with}\quad
    {\cal M}_{\rm inf} = {\cal M}_{n,2}(1',2',\dots,n') \label{inf3} \, ,
\end{equation}
which is the same form as for the $n$-line and $(n{-}1)$-line cases. This is because the only problematic bracket $[12]$ never shows up explicitly in the Hodges formula. Note that this analysis does not depend on the helicities of the unshifted legs and all three cases (\ref{exthel}) are here unified. The set of spinors (\ref{prime4}) is interesting because $\widetilde{\lambda}_{1'}$ and $\widetilde{\lambda}_{2'}$ are collinear due to
\begin{equation}
    [1'2']=0\, ,
\end{equation}
which is an antiholomorphic collinear limit. If this was a factorization pole of the amplitude, the result would be a simple factorization into two amplitudes:
\begin{equation}
	\begin{tabular}{cc}
	 \includegraphics[scale=.72]{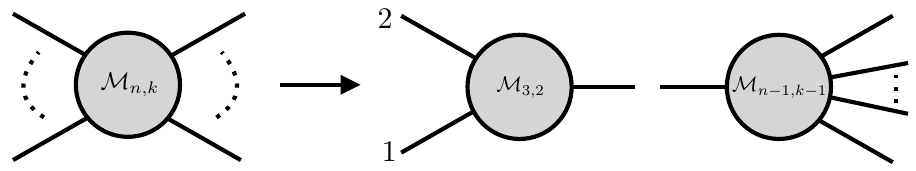} \label{fact0}
	\end{tabular} 
\end{equation}
However, this can not happen here as $k=2$ and the amplitude on the right side would be $k=1$, and hence vanishing for $n>3$. Interestingly, the same argument applies for $(n{-}3)$-line shift. Using the same parametrization of the Hodges formula, we never encounter problematic square brackets $[12]$, $[13]$, $[23]$ in $\Phi_{a,b}$ the leading term at infinity is given by the same expression (\ref{inf3}) where now 
\begin{equation}
     \widetilde{\lambda}_{a'} \equiv \left(\begin{matrix}0 \\ [a\tilde{\eta}] \end{matrix}\right)\,\,\, \mbox{for}\,\,a=1,2,3,\,\,\,\, \mbox{and}
     \,\,\,\,\,\,\widetilde{\lambda}_{a'} \equiv \left(\begin{matrix}\alpha_a \\ [a\tilde{\eta}] \end{matrix}\right)\,\,\,\mbox{for others}\label{prime5}
\end{equation}
For $(n{-}4)$ and lower line shifts we can not make the same argument. There is no way to avoid the presence of square brackets between two unshifted legs, and in the limit $z\rightarrow\infty$ we get a mix of $z^1$ and $z^0$ terms. Also, other representations of the MHV amplitude, such as BCFW recursion relations \cite{Elvang:2007sg,Drummond:2009ge}, BGK \cite{Berends:1988zp}, Mason-Skinner \cite{Mason:2009afn} or the inverse-soft factor formulas \cite{Bern:1998sv,Nguyen:2009jk}, do not make the result manifest either.  

We checked numerically for $(n{-}p)$ line shift for $p\geq4$ that the amplitude exhibits precisely the same behavior,
\begin{equation}
{\cal M}_{n,2} = z^{n-3} \times {\cal M}_{\rm inf} + {\cal O}(z^{n-4})\quad\mbox{with}\quad
    {\cal M}_{\rm inf} = {\cal M}_{n,2}(1',2',\dots,n') \label{inf4}
\end{equation}
The scaling is consistent with the all-line shift (\ref{inf1}) and it seems that $\sim z^{n-3}$ does not depend on a particular shift. For $\lambda$ spinors we always have $\lambda_{a'}=\lambda_a$, while for $p$ unshifted and $n{-}p$ shifted $\widetilde{\lambda}$ spinors we have,
\begin{equation}
     \widetilde{\lambda}_{a'} \equiv \left(\begin{matrix}0 \\ [a\tilde{\eta}] \end{matrix}\right)\,\,\, \mbox{for}\,\,a=1{\dots},p,\,\,\,\,\,\,\,\,\widetilde{\lambda}_{a'} \equiv \left(\begin{matrix}\alpha_a \\ [a\tilde{\eta}] \end{matrix}\right)\,\,\, \mbox{for}\,\,a=p{+}1,{\dots},n\label{prime6}
\end{equation}
where both type of spinors have the same, the unshifted have $\alpha_a=0$. We checked cases $p=4,5,6,7$ for up to $n=10$ each. Hence we conjecture that the formula (\ref{inf4}) is valid for any $p\geq2$ and any $n$. Kinematics (\ref{prime6}) makes all $\widetilde{\lambda}_{a'}$ for $a=1,{\dots},p$ proportional which is an anti-holomorphic collinear limit. This is very singular configuration, and if this was a sequence of factorization poles it would give a non-zero result only if $k{-}p+1\geq2$, which is not satisfied for the MHV case.

\subsection{$\overline{\rm MHV}$ amplitudes}

Before discussing general $k$, we first focus on the case of $\overline{\rm MHV}$ amplitudes.We can conveniently write them using a parity conjugation of the Hodges formula for MHV amplitudes,
\begin{equation}
    {\cal M}_{n,n{-}2}(1^+,2^+,3^-,{\dots},n^-) = [12]^8\times\frac{{\rm det}(\Psi_{abc}^{def})}{[abc][def]} \label{Hodges1}
\end{equation}
where now $[abc]=[ab][bc][ca]$, and $\Psi_{abc}^{def}$ is the $(n{-}3\,\times\,n{-}3)$ matrix obtained from the $n\times n$ matrix $\Psi$ by deleting rows $a,b,c$ and columns $d,e,f$, analogous to the $\Phi$ matrix. The elements of the $\Psi$ matrix are defined as
\begin{equation}
    \Psi_{i,j} = \frac{\la ij\ra }{[ij]},\qquad \Psi_{i,i} = - \sum_{j\neq i}\frac{\la ij\ra[ja][jb]}{[ij][ia][ib]} \label{Hodges3}
\end{equation}
where $\widetilde{\lambda}_a$, $\widetilde{\lambda}_b$ are arbitrary spinors. The behavior of the $\overline{\rm MHV}$ amplitude at infinity is very different than for the MHV amplitude, as the shift is anti-holomorphic and breaks the parity symmetry. 

To see the result most transparently, we need to choose a very specific parametrization of the $\Psi$ matrix, namely $(a,b,c)=(1,3,4)$ and $(d,e,f)=(2,3,4)$. The helicity factor will eventually be important, but we will factor that in at the end after we analyze the skeleton (helicity insensitive) amplitude,
\begin{equation}
    {\cal M}_{n,n{-}2}(1,2,\widehat{3},\widehat{4},\dots,\widehat{n}) = \frac{[ij]^8}{[1\widehat{3}][1\widehat{4}][2\widehat{3}][2\widehat{4}][\widehat{3}\widehat{4}]^2} \times {\rm det}(\Psi_{134}^{234}) \label{Hodges3}
\end{equation}
The $\Psi$ matrix for this choice takes the form,
\begin{equation}
    \Psi_{134}^{234} = \left(\begin{matrix}\Psi_{1,2} & \Psi_{1,5} & \Psi_{1,6} & \dots & \Psi_{1,n}\\ \Psi_{2,5} & \Psi_{5,5} & \Psi_{5,6} & \dots & \Psi_{5,n} \\ \Psi_{2,6} & \Psi_{5,6} & \Psi_{6,6} & \dots & \Psi_{6,n} \\
    \vdots & \vdots & \vdots & \ddots & \vdots \\ \Psi_{2,n} & \Psi_{5,n} & \Psi_{6,n} & \dots & \Psi_{n,n}\end{matrix}\right)
\end{equation}
For $z\rightarrow\infty$ the individual elements scale as
\begin{equation}
    \Psi_{1,2} = \frac{\la 12\ra}{[12]}, \,\,\,\, \Psi_{1,a} = \frac{\la 1a\ra}{[1\widehat{a}]} \rightarrow \frac{1}{z}\frac{\la 1a\ra}{[1a']}, \,\,\,\,\Psi_{2,a} = \frac{\la 2a\ra}{[2\widehat{a}]} \rightarrow \frac{1}{z}\frac{\la 2a\ra}{[2a']}, \,\,\,\, \Psi_{a,b} = \frac{\la ab\ra}{[\hat{a}\hat{b}]} \rightarrow \frac{1}{z}\frac{\la ab\ra}{[a'b']},
\end{equation}
for shifted labels $a,b$. We already wrote the result in terms of the prime spinors defined in (\ref{prime1}). Looking more closely at $\Psi$ matrix, in the first row all elements scale as ${\sim z^{-1}}$ except the element $\Psi_{1,2}$ which scales as ${\sim} z^0$, and hence this term dominates. As a result, we can ignore all $\Psi_{1,k}$ terms for $k>2$, and rewrite the determinant at the leading order at large $z$ as
\begin{equation}
    {\rm det}(\Psi_{134}^{234}) = \Psi_{1,2} \times {\rm det}(\widetilde{\Psi})
\end{equation}
where 
\begin{equation}
    \widetilde{\Psi} = \left(\begin{matrix} \Psi_{5,5} & \Psi_{5,6} & \dots & \Psi_{5,n} \\\Psi_{5,6} & \Psi_{6,6} & \dots & \Psi_{6,n} \\
    \vdots & \vdots & \ddots & \vdots \\ \Psi_{5,n} & \Psi_{6,n} & \dots & \Psi_{n,n}\end{matrix}\right)
\end{equation}
The non-diagonal elements of this matrix are just given by 
\begin{equation}
    \Psi_{a,b} = \frac{1}{z}\frac{\la ab\ra}{[a'b']}
\end{equation}
The diagonal elements are more interesting. Let us look at $\Psi_{5,5}$,
\begin{equation}
    \Psi_{5,5} = \sum_{a=1}^2 \underbrace{\frac{\la 5a\ra[ax][ay]}{[\hat{5}a][\hat{5}x][\hat{5}y]}}_{1/z^3} + \sum_{b=3}^n \underbrace{\frac{\la 5b\ra[\hat{b}x][\hat{b}y]}{[\hat{5}\hat{b}][\hat{5}x][\hat{5}y]}}_{{1/z}}\rightarrow \frac{1}{z}  \sum_{b=3}^n \frac{\la 5b\ra\alpha_b^2}{[5'b']\alpha_5^2}
\end{equation}
where the sum excludes $b=5$. This means that the terms with labels $1,2$ dropped from the sum as subleading for large $z$. 
The last factor looks exactly like a five-point MHV amplitude! In fact, we can define:
\begin{equation}
    \lambda_\omega \equiv (1{+}2)|\widetilde{\eta}],\quad \widetilde{\lambda}_\omega \equiv \left(\begin{array}{c}0\\1\end{array}\right) \label{spinors}
\end{equation}
for which
\begin{equation}
    \alpha_a = [a'\omega]
\end{equation}
Using that we can rewrite the $\Psi_{5,5}$ element as 
\begin{equation}
    \Psi_{5,5} = \sum_{b=3}^n \frac{\la 5b\ra[b'\omega]^2}{[5'b'][5\omega]^2}
\end{equation}
which can be also shown to be equal to (if we reintroduce auxiliary spinors $\widetilde{\lambda}_x$, $\widetilde{\lambda}_y$),
\begin{equation}
    \Psi_{5,5} = \sum_{b=3}^n \frac{\la 5b\ra[b'x][b'y]}{[5'b'][5x][5y]} + \frac{\la 5\omega\ra[\omega x][\omega y]}{[5'\omega][5'x][5'y]}
\end{equation}
We can also check that our newly defined spinors $\lambda_\omega$, $\widetilde{\lambda}_{\omega}$ together with the prime spinors $\lambda_{a'}$, $\widetilde{\lambda}_{a'}$ satisfy momentum conservation,
\begin{equation}
    \sum_{a=3}^n \lambda_{a'}\widetilde{\lambda}_{a'} +\lambda_\omega\widetilde{\lambda}_\omega = 0
\end{equation}
The matrix $\widetilde{\Psi}$ represents an honest Hodges matrix for $n{-}1$ labels $\omega,3',4',5',6',\dots,n'$,
\begin{equation}
    (\widetilde{\Psi})_{{n{-}3}} = \left(\Psi_{\omega\,3'4'}^{\omega\,3'4'}\right)_{n{-}4}
\end{equation}
where we put the labels to denote that the matrix on the left hand side is $(n{-}3)\times(n{-}3)$ while the matrix on the right is $(n{-}4)\times(n{-}4)$. For the determinant ${\rm det}(\Psi_{134}^{234})$ we get,
\begin{equation}
      {\rm det}(\Psi_{134}^{234}) = \frac{1}{z^{n{-}4}}\times \frac{\la12\ra}{[12]} \times {\rm det}\left(\Psi_{\omega\,3'4'}^{\omega\,3'4'}\right)
\end{equation}
Finally, the prefactor can be now rewritten as
\begin{equation}
    \frac{1}{[1\widehat{3}][1\widehat{4}][2\widehat{3}][2\widehat{4}][\widehat{3}\widehat{4}]^2} = \frac{1}{z^6}\times \frac{1}{[1\widetilde{\eta}]^2[2\widetilde{\eta}]^2}\times \frac{1}{[3'4']^2[3'\omega]^2[4'\omega]^2}
\end{equation}
Putting everything together, we can conclude that 
\begin{equation}
    {\cal M}_{n,n{-}2}(1,2,\hat{3},\hat{4},{\dots},\hat{n}) = \frac{1}{z^{n+2}}\times {\cal M}_{\rm inf} + {\cal O}\left(\frac{1}{z^{n{+}3}}\right)\label{MHVbar1}
\end{equation}
where
\begin{equation}
    {\cal M}_{\rm inf}= s_{12}\times  \frac{H^8}{[12]^2[1\widetilde{\eta}]^2[2\widetilde{\eta}]^2}\times \widetilde{\cal M}_{n{-}1,n{-}3}(\omega,3',4',{\dots},n') \label{MHVbar2}
\end{equation}
where $H$ is the helicity factor (which depends on $z$). Note that $\widetilde{\cal M}_{n{-}1,n{-}3}$ is an honest $(n{-}1)$-point $\overline{\rm MHV}$ amplitude without a helicity factor. We can also recognize the form of the three-point $\overline{\rm MHV}$ amplitude in (\ref{MHVbar2}) for labels $1,2,\eta$ such that $\widetilde{\lambda}_\eta = \widetilde{\eta}$. Hence, the expression (\ref{MHVbar1}) resembles a \emph{factorization}, 
\begin{equation}
	\begin{tabular}{cc}
	 \includegraphics[scale=.72]{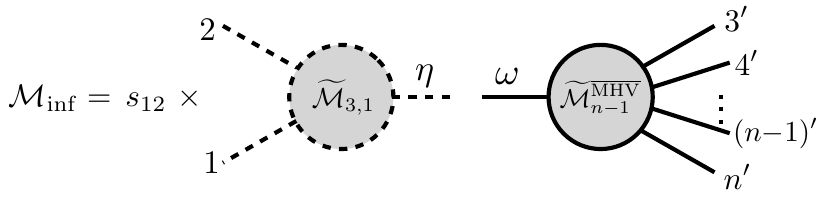} \label{fact1}
	\end{tabular} 
\end{equation}
However, $\widetilde{\cal M}_{3,1}$ is not an honest amplitude for momenta given by spinors $\lambda_1$, $\widetilde{\lambda}_1$, $\lambda_2$, $\widetilde{\lambda}_2$, and any choice for $\eta$ kinematics. If it was the honest amplitude, then $\la 12\ra=0$ due to the three-point kinematics, but this is not the case, hence we label it in the figure (\ref{fact1}) with a dashed line. 

Finally, we have to resolve the problem of the helicity factor, which would be now sensitive to helicities of unshifted legs $1,2$. It breaks down to three cases (\ref{exthel}), for which the helicity factor can be rewritten as
\begin{align}
{\cal M}_{n,n{-}2}(\{2,0\}) \,:\,\,\,& H= z^8\times [n{-}1\,n]^8\\
{\cal M}_{n,n{-}2}(\{1,1\}) \,:\,\,\,& H= z^8\times [2\widetilde{\eta}]^8[\omega n]^8\\
{\cal M}_{n,n{-}2}(\{0,2\}) \,:\,\,\,& H= [12]^8
\end{align}
This gives us three cases 
\begin{align}
   {\cal M}_{n,n{-}2}(\{2,0\}) &= (-1)^n \frac{s_{12}}{z^{n-6}}\times \frac{1}{[12]^2[1\widetilde{\eta}]^2[2\widetilde{\eta}]^2}\times {\cal M}_{n{-}1,n{-}3}(\omega^-,3'^-,{\dots},(n{-}1)'^+,n'^+) +{\cal O}(z^{5-n})\nonumber\\
  {\cal M}_{n,n{-}2}(\{1,1\}) &= (-1)^n\frac{s_{12}}{z^{n-6}}\times \frac{[2\widetilde{\eta}]^8}{[12]^2[1\widetilde{\eta}]^2[2\widetilde{\eta}]^2}\times {\cal M}_{n{-}1,n{-}3}(\omega^+,3'^-,{\dots},(n{-}1)'^-,n'^+)+{\cal O}(z^{5-n})\nonumber\\
  {\cal M}_{n,n{-}2}(\{0,2\}) &=(-1)^n \frac{s_{12}}{z^{n+2}}\times \frac{[12]^8}{[12]^2[1\widetilde{\eta}]^2[2\widetilde{\eta}]^2}\times \widetilde{\cal M}_{n{-}1,n{-}3}(\omega,3',{\dots},(n{-}1)',n')+{\cal O}(z^{-3-n}) \label{MHVbar3}
\end{align}
where the scaling is worse at infinity than for the all line shift (\ref{inf1}), which gives $\sim z^{5-n}$ for $\overline{\rm MHV}$ amplitudes. For $\{2,0\}$ and $\{1,1\}$ configurations the helicity factor $H$ produces the right factors for the amplitude $\widetilde{\cal M}_{n{-}1,n{-}3}$ turning it into a proper amplitude ${\cal M}_{n{-}1,n{-}3}$ with assigned helicities for all legs including $\omega$. Furthermore, for $\{1,1\}$ we also get the right helicity factor for the three-point $\overline{\rm MHV}$ quasi-amplitude, which would correspond to ${\cal M}_{3,1}(1^-,2^+,\eta^+)$. However, as mentioned earlier, this is not on the support of the proper three-point kinematics. For the configuration $\{0,2\}$ the helicity factor $H$ does not provide the right terms for either of the expressions. Note that the extra factor $s_{12}$ in the formula (\ref{fact1}) also indicates that the pattern at infinity is differs from the residues on factorization poles.

\subsection{Six-point $\rm NMHV$ amplitudes}

Beyond MHV we do not have an analogue of the compact Hodges formula, and though there are explicit formulas for higher point N$^k$MHV amplitudes in the literature, we proceed here case by case and then do numerical checks for higher $n$ to prove our conjectures. The first case of interest is the six-point NMHV amplitude. In this case the result is very sensitive to the helicities of unshifted legs. Three different cases (\ref{exthel}) are
\begin{equation}
	\begin{tabular}{cc}
\hspace{-0.6cm}  \includegraphics[scale=.75]{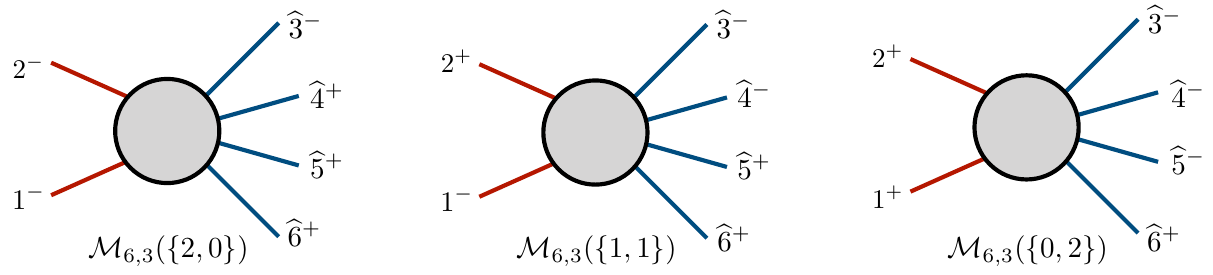}
	\end{tabular} 
\end{equation}
Let us start with the first case, perform the the multi-line anti-holomorphic shift (\ref{shift}) and take the limit $z\rightarrow\infty$. As a result, we get
\begin{equation}
        {\cal M}_{6,3}(\{2,0\}) = z^2 {\cal M}_{\rm inf}(\{2,0\}) + {\cal O}(z)
\end{equation}
where we denoted the leading amplitude at infinity as
\begin{equation}
    {\cal M}_{\rm inf}(\{2,0\})= \frac{\la 3|1{+}2|\widetilde{\eta}]^8\la12\ra
    \Bigg\{
    \begin{array}{c}
    (\alpha_3[4\widetilde{\eta}]-\alpha_4[3\widetilde{\eta}])(\alpha_5[6\widetilde{\eta}]-\alpha_6[5\widetilde{\eta}])\la35\ra\la46\ra\\
    -(\alpha_3[5\widetilde{\eta}]-\alpha_5[3\widetilde{\eta}])(\alpha_4[6\widetilde{\eta}]-\alpha_6[4\widetilde{\eta}])\la34\ra\la56\ra
    \end{array}
    \Bigg\}}{[12][1\widetilde{\eta}]^2[2\widetilde{\eta}]^2\la34\ra\la35\ra\la36\ra\la45\ra\la46\ra\la56\ra\la3|1{+}2|\widetilde{\eta}]\la4|1{+}2|\widetilde{\eta}]\la5|1{+}2|\widetilde{\eta}]\la6|1{+}2|\widetilde{\eta}]} \label{InfNMHV1}
\end{equation}
which we can recognize upon inspection and using the prime spinors (\ref{prime1}) and (\ref{spinors}), as being equal to
\begin{align}
 {\cal M}_{\rm inf}(\{2,0\}) &= s_{12} \times \frac{1}{[12]^2[1\widetilde{\eta}]^2[2\widetilde{\eta}]^2}\times \la3\omega\ra^8\times \widetilde{\cal M}_{5,2}(\omega,3',4',5',6')\nonumber\\
 &= s_{12} \times \frac{1}{[12]^2[1\widetilde{\eta}]^2[2\widetilde{\eta}]^2}\times  {\cal M}_{5,2}(\omega^-,3'^-,4'^+,5'^+,6'^+)\label{InfNMHV2}
\end{align}
This is the same form as we got for $\overline{\rm MHV}$ amplitudes (\ref{MHVbar3}), and it again resembles a factorization pattern.  
\begin{equation}
	\begin{tabular}{cc}
\includegraphics[scale=.7]{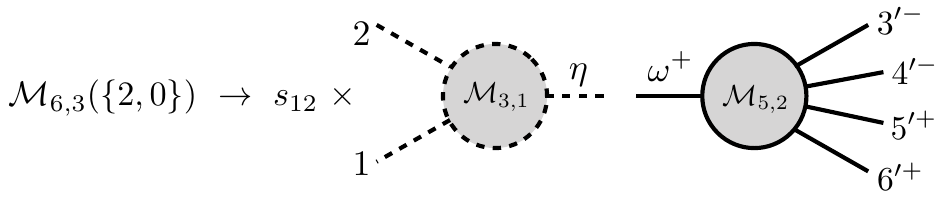}
	\end{tabular} 
\end{equation}
through the same argument about general kinematics $[12]\neq 0$ applies here. The next amplitude is ${\cal M}_{6,3}(\{1,1\})$, and the analogous computation gives us
\begin{equation}
        {\cal M}_{6,3}(\{1,1\}) = z^2 {\cal M}_{\rm inf}(\{1,1\}) + {\cal O}(z)
\end{equation}
which is the same scaling as for the amplitude ${\cal M}_{6,3}(\{2,0\})$, but now
\begin{equation}
    {\cal M}_{\rm inf}(\{1,1\}) = \frac{[2\widetilde{\eta}]^8\la34\ra^8s_{12}
    \Bigg\{
    \begin{array}{c}
    (\alpha_3[6\widetilde{\eta}]-\alpha_6[3\widetilde{\eta}])(\alpha_4[5\widetilde{\eta}]-\alpha_5[4\widetilde{\eta}])\la34\ra\la56\ra\\
    -(\alpha_3[4\widetilde{\eta}]-\alpha_4[3\widetilde{\eta}])(\alpha_5[6\widetilde{\eta}]-\alpha_6[5\widetilde{\eta}])\la36\ra\la45\ra
    \end{array}
    \Bigg\}}{[1\widetilde{\eta}]^2[2\widetilde{\eta}]^2[12]^2\la34\ra\la35\ra\la36\ra\la45\ra\la46\ra\la56\ra\la3|1{+}2|\widetilde{\eta}]\la4|1{+}2|\widetilde{\eta}]\la5|1{+}2|\widetilde{\eta}]\la6|1{+}2|\widetilde{\eta}]} \label{InfNMHV3}
\end{equation}
but using the new spinors (\ref{prime1}) and (\ref{spinors}) we immediately identify 
\begin{equation}
 {\cal M}_{\rm inf}(\{1,1\}) = s_{12} \times \frac{[2\widetilde{\eta}]^8}{[12]^2[1\widetilde{\eta}]^2[2\widetilde{\eta}]^2}\times{\cal M}_{5,2}(\omega^+,3'^-,4'^-,5'^+,6'^+) \label{InfNMHV4}
\end{equation}
where the second factor has the correct form of the amplitude ${\cal M}_{3,1}(1^-,2^+,\eta^+)$
\begin{equation}
	\begin{tabular}{cc}
\includegraphics[scale=.7]{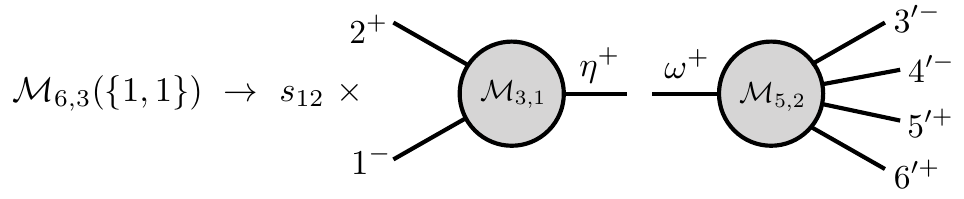}
	\end{tabular} 
\end{equation}
For the last helicity amplitude, ${\cal M}_{6,3}(\{0,2\})$, the behavior at infinity is suppressed in comparison to ${\cal M}_{6,3}(\{2,0\})$ and ${\cal M}_{6,3}(\{1,1\})$,
\begin{equation}
        {\cal M}_{6,3}(\{0,2\}) = {\cal O}(z^1)
\end{equation}
Upon inspection the coefficient of the ${\sim}z^1$ term can not be recognized as something nice, and it can be even proven that no factorization of indices happens in this case. Note that the scaling of all cases of ${\cal M}_{6,3}$ under the all-line shift would be ${\cal O}(z)$, hence ${\cal M}_{6,3}(\{2,0\})$ and ${\cal M}_{6,3}(\{1,1\})$ have even stronger scaling at infinity, while ${\cal M}_{6,3}(\{0,2\})$ exhibits only a subleading behavior.

\subsection{Leading amplitude at infinity for all $n$ and $k$}

The pattern observed for the six-point NMHV amplitude generalizes for $n>6$ and for $k=(3,4,\dots,n{-}3)$, ie. all cases except MHV and $\overline{\rm MHV}$ we studied separately earlier. For the three helicity patterns ${\cal M}_{n,k}(\{2,0\})$, ${\cal M}_{n,k}(\{1,1\})$, ${\cal M}_{n,k}(\{0,2\})$,
\begin{equation}
	\begin{tabular}{cc}
	 \includegraphics[scale=.7]{fig10.pdf}
	\end{tabular} \nonumber
\end{equation}
we have a direct generalization of formulas from the previous subsections, namely
\begin{align}
        {\cal M}_{n,k}(\{2,0\}) &= z^{n-2k+2}\times {\cal M}_{\rm inf}(\{2,0\}) + {\cal O}(z^{n-2k+1}) \label{res1}\\
        {\cal M}_{n,k}(\{1,1\}) &= z^{n-2k+2}\times {\cal M}_{\rm inf}(\{1,1\}) + {\cal O}(z^{n-2k+1}) \label{res2}\\
        {\cal M}_{n,k}(\{0,2\}) &= {\cal O}(z^{n-2k+1}) \label{res3}
\end{align}
where the first two configurations exhibit the same factorization pattern, up to a possible overall sign depending on helicity sector and multiplicity,
\begin{align}
 {\cal M}_{\rm inf}(\{2,0\})  &= s_{12} \times \frac{1}{[12]^2[1\widetilde{\eta}]^2[2\widetilde{\eta}]^2}\times  {\cal M}_{n{-}1,k{-}1}(\omega^-,3'^-,{\dots},k'^-,(k{+}1)'^+,{\dots},n'^+)\nonumber\\
 {\cal M}_{\rm inf}(\{1,1\})  &= s_{12} \times \frac{[2\widetilde{\eta}]^8}{[12]^2[1\widetilde{\eta}]^2[2\widetilde{\eta}]^2}\times  {\cal M}_{n{-}1,k{-}1}(\omega^+,3'^-,{\dots},(k{+}1)'^-,(k{+}2)'^+,{\dots},n'^+) \label{Infgen}
\end{align}
Note that the same behavior also applies to $\overline{\rm MHV}$ amplitudes for the $(\{2,0\})$ and $(\{1,1\})$ cases (\ref{MHVbar3}). The exception is the $(\{0,2\})$ configuration which has accidentally the factorization structure for the $k=n{-}2$, $\overline{\rm MHV}$ case but this does not generalize beyond. We can interpret this failure as an inconsistency between the labels of the original amplitude ${\cal M}_{n,k}(\{0,2\})$ and the factorization form, namely the amplitude ${\cal M}_{n{-}1,k{-}1}$ does not have a proper helicity assignment for the new leg $\omega$ such that the helicity works out,
\begin{equation}
   {\cal M}_{n{-}1,k{-}1}(\omega^{\ast},3'^-,{\dots},(k{+}2)'^-,(k{+}3)'^+,{\dots},n'^+) 
\end{equation}
because whatever helicity assignment we make for $\omega$, the total helicity degree will be at least $k$, and not $k{-}1$. The exception of the $\overline{\rm MHV}$ case comes from the fact that the amplitude depends on a particular assignment of helicities only through the helicity factor. But note that in this case the scaling is further enhanced by $z^8$, so it should be considered as special case. 

From the point of view of scaling (\ref{Infgen}), the MHV amplitudes also fit to our story, as they scale one power better than predicted $\sim z^{n-2}\rightarrow z^{n-3}$. This can be again understood by the non-existence of the ${\cal M}_{n-1,k-1}$ amplitude on the right hand side. In this case, it is just a simple statement that $k-1=1$ for MHV case, and the amplitude ${\cal M}_{n-1,1}$ just vanishes identically. That is also the reason why we did not observe any factorization there. Hence we can conclude that (\ref{Infgen}) describes correctly the leading behavior at infinity for all $n,k$. 

\subsection{Subleading amplitude at infinity}

The MHV analysis suggests some information about the subleading term of order $\sim z^{n-2k+1}$, which in fact is a leading term for the MHV case. We showed that the coefficient of the $z^{n-3}$ term is the same amplitude ${\cal M}_{n,2}$ evaluated on special kinematics (\ref{inf4}), where notably 
two of the spinors $\widetilde{\lambda}_{1'}$, $\widetilde{\lambda}_{2'}$ were in the collinear configuration $[1'2']=0$ while other $\widetilde{\lambda}_{a'}$ spinors took the same form as in all other cases. At the same time there was no new particle $\omega$.

We can explore this possibility for higher $n,k$ starting from the cases for which $\sim z^{n-2k+1}$ is dominant as the $z^{n-2k+2}$ leading term vanishes. These are classes of ${\cal M}_{n,k}(\{0,2\})$ amplitudes. The proposal is that 
\begin{equation}
    {\cal M}_{n,k}(\{0,2\}) = z^{n-2k+1}\times {\cal M}_{n,k}(1',2',\dots,n') + {\cal O}(z^{n-2k}) \label{subleading}
\end{equation}
where the amplitude on the right is evaluated on the same special kinematics (\ref{inf4}), namely, 
\begin{equation}
     \widetilde{\lambda}_{1'} \equiv \left(\begin{matrix}0 \\ [1\tilde{\eta}] \end{matrix}\right),\quad   \widetilde{\lambda}_{2'} \equiv \left(\begin{matrix}0 \\ [2\tilde{\eta}] \end{matrix}\right),\quad
     \widetilde{\lambda}_{a'} \equiv \left(\begin{matrix}\alpha_a \\ [a\tilde{\eta}] \end{matrix}\right)
\end{equation}
As discussed before, this is a collinear kinematics in $\widetilde{\lambda}_{1'},\widetilde{\lambda}_{2'}$, ie. $[1'2']=0$. Note that this expression does not appear as the pole of ${\cal M}_{n,k}(\{0,2\})$ with prime spinors, because of the corresponding factorization channel -- which requires to factor out three-point $\overline{\rm MHV}$ amplitude ${\cal M}_{3,1}$. This is inconsistent with the positive helicities of particles $1$ and $2$,
\begin{equation}
	\begin{tabular}{cc}
	 \includegraphics[scale=.7]{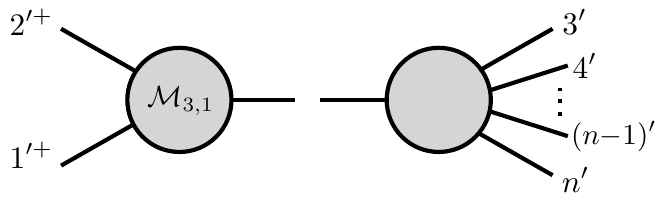}
	\end{tabular} \nonumber
\end{equation}
We checked numerically this proposal for $k=3,4,5$ up to $n=9$ and the formula (\ref{subleading}) indeed holds. In Table 1 is the summary of numerical checks: the triplet of numbers $\{a,b,c\}$ gives the scaling for $\{\{2,0\} , \{1,1\},\{0,2\}\}$ configurations of ${\cal M}_{n,k}$,
    \begin{table}
     \[
    \begin{array}{c|cccc}
    (n,k) & n=6&n=7&n=8&n=9\\
    \hline 
    k =2 & \{ z^3,z^3,z^3 \} & \{ z^4,z^4,z^4 \} & \{z^5,z^5,z^5\} & \{z^6,z^6,z^6\}  \\
    k =3 & \{ {\color{red} z^2},{\color{red}z^2},{z^1} \}& \{ {\color{red}z^3},{\color{red}z^3},{z^2} \} & \{ {\color{red} z^4},{\color{red} z^4},z^3\} & \{  {\color{red} z^5},{\color{red} z^5},{z^4}   \}   \\
    k = 4 &\{ {\color{red} z^0},{\color{red}z^0},{\color{red} z^{-8}} \} & \{ {\color{red}z^1},{\color{red}z^1},{z^0} \} & \{ {\color{red}z^2},{\color{red}z^2},{z} \} & \{ {\color{red} z^3},{\color{red} z^3},{z^2} \} \\  
    k = 5 & {\rm -}& \{ {\color{red} z^{-1}},{\color{red} z^{-1}},{ \color{red}z^{-9}} \}& \{ {\color{red} z^0 },{\color{red} z^0 },{z^{-1}} \}& \{ {\color{red}z^1},{\color{red}z^1},{z^0} \}\\
    k = 6 & {\rm -} & {\rm -} & \{ {\color{red}z^{-2}},{\color{red}z^{-2}},{\color{red}z^{-10}} \} & \{ {\color{red}z^{-1}},{\color{red}z^{-1}},{z^{-2}} \}  \\
    k = 7 & {\rm -} & {\rm -} & {\rm -} & \{ {\color{red}z^{-3}},{\color{red}z^{-3}},{\color{red}z^{-11}} \}
    \end{array}
    \]
    \caption{Scaling at infinity for $(n{-}2)$ shift. The red color denotes the amplitudes that manifest factorization at infinity, while the black ink denotes cases where the amplitude satisfies (\ref{subleading}). The dash means that the corresponding amplitude does not exist. }
    \end{table}
You can clearly see the correlation between the pattern and the $z$-scaling, supporting our conjecture. The only exception is the $\overline{\rm MHV}$ amplitude ${\cal M}_{n,n{-}2}(\{0,2\})$ that has a different scaling because of the helicity factor. There is a natural question if the subleading ${\cal O}(z^{n-2k+1})$ terms in (\ref{res1}) and (\ref{res2}) also obey the same (\ref{subleading}) behavior.
\begin{equation}
        {\cal M}_{n,k}(\{2,0\}) = z^{n-2k+2}\times {\cal M}_{\rm inf}(\{2,0\}) + z^{n-2k+1}\times{\cal M}_{\rm sub}(\{2,0\}) +  {\cal O}(z^{n-2k})
\end{equation}
where ${\cal M}_{\rm sub} = {\cal M}_{n,k}(1',2',\dots,n')$ for the same prime spinors (\ref{inf4}). However, this can not be the case as ${\cal M}_{\rm sub}$ clearly blows up for $[1'2']=0$, and the term is divergent. If the subleading term can be obtained as the same amplitude ${\cal M}_{n,k}(\{2,0\})$ on yet another special spinors (which do not make the amplitude to blow up), is an open question.

\section{Generalizations}

In this section, we explore certain generalizations of our analysis, including the $(n-n_1)$-line shifts for $n_1>2$, and the comparison between behavior at infinity between gravitons and gluons.

\subsection*{Scaling at infinity for $(n{-}n_1)$ line shifts}

For the $(n{-}3)$-line shift, we sort the helicity sector choices as: $\{ \{ 3,0 \} ,\{ 2,1 \},\{ 1,2 \},\{ 0,3 \}\} $, denoting the number of $\{-,+\}$ unshifted legs, and provide the scalings at infinity for various $n,k$ in Table 2.
    \begin{table}
     \[
    \begin{array}{c|cccc}
    (n,k) & n=6&n=7&n=8&n=9\\
    \hline 
    k =2 & \{ {\rm -},{z^3},{z^3},{z^3}\}  & \{ {\rm -},{z^4},{z^4},{z^4}\} & \{ {\rm -},{z^5},{z^5},{z^5}\} & \{ {\rm -},{z^6},{z^6},{z^6}\}\\
    k =3 &   \{ {z^2},{z^2},{z^2},{z^{-6}}\} &  \{ {z^3},{z^3},{z^3},{z^2}\}  &  \{ {z^4},{z^4},{z^4},{z^3}\}  &  \{ {z^5},{z^5},{z^5},{z^4}\}  \\
    k = 4 &  \{ {z^1},{z^1},{z^{-7}},{\rm -} \}   &  \{ {z^2},{z^2},{z^{1}},{z^{-7}} \} &  \{ {z^3},{z^3},{z^{2}},{z^{1}} \}  & \{ {z^4},{z^4},{z^{3}},{z^{2}} \}  \\
    k = 5 & {\rm -}& \{ {z^0},{z^0},{z^{-8} },{\rm -} \} & \{ {z^1},{z^1},{z^{0}},{z^{-8}}  \}  & \{ {z^2},{z^2},{z^{1}},{z^{0}}  \}  \\
    k = 6 & {\rm -} & {\rm -}  & \{  {z^{-1}},{z^{-1}},{z^{-9}}  ,{\rm -}   \} & \{  {z^0},{z^0},{z^{-1}},{z^{-9}}   \} \\
    k = 7 & {\rm -} & {\rm -} & {\rm -} & \{  {z^{-2}},{z^{-2}}, {  z^{-10} }, {\rm -} \}
    \end{array}
    \]
    \caption{Scaling behavior for $(n{-}3)$-line shift. }
    \end{table}
These data suggest a following scaling
\begin{align}
    {\cal M}_{n,k}(\{3,0\})/ {\cal M}_{n,k}(\{2,1\})&\sim z^{n-2k+3} \qquad \mbox{for}\,\,\, k>3\\
    {\cal M}_{n,k}(\{1,2\}) &\sim z^{n-2k+2} \qquad \mbox{for}\,\,\, n{-}2>k>2\\
    {\cal M}_{n,k}(\{0,3\}) &\sim z^{n-2k+1} \qquad \mbox{for}\,\,\, k<n{-}3
\end{align}
All shift configurations of MHV amplitudes scale like $\sim z^{n{-}3}$. For $\{3,0\}$ and $\{2,2\}$ configurations the NMHV amplitude has $\sim z^{n{-}4}$ scaling.

Comparison of both tables suggests that for a general $(n{-}n_1)$-line shift, at least for $\{n_1,0\}$ and $\{n_1{-}1,1\}$ configurations we have 
\begin{equation}
    {\cal M}_{n,k} \sim z^{n-2k+n_1} \,\,\,\mbox{for}\,\,\, k>n_1 \label{scale1}
\end{equation}
while the $\{0,n_1\}$ configuration exhibits
\begin{equation}
    {\cal M}_{n,k} \sim z^{n-2k+1} \,\,\,\mbox{for}\,\,\, n{-}n_1>k\label{scale2}
\end{equation}
We checked for $n=8,9$ data for the $(n{-}4)$ line shift which confirms both formulas (\ref{scale1}), (\ref{scale2}), and further suggests an extension to arbitrary helicity configuration $\{a,n_1{-}a\}$,
\begin{equation}
    {\cal M}_{n,k} \sim z^{n-2k+a+1} \,\,\,\,\,\mbox{for}\,\,\,\,\, n{-}n_1{+}a>k>a{+}1 \label{scale3}
\end{equation}  
for $a<n_1$. All data up to $n=9$ are consistent with (\ref{scale3}).

\subsection*{Comparison to other work}

In \cite{Edison:2019ovj}, the scaling of ${\cal M}_{n,k}$ under a shift of $m\leq k$ negative helicity gravitons was considered. In our notation, this is configuration $\{k{-}m,n{-}k\}$ and the scaling for this shift was conjectured to be 
\begin{equation}
    {\cal M}_{n,k}(z) \sim z^{n-k-6-m} \quad\mbox{for}\,\,\,\,z\rightarrow\infty
\end{equation}
For $(n{-}2)$-line shift this has overlap only with $\overline{\rm MHV}$ amplitude for $\{0,2\}$ configuration. In this notation, this is $k=m=n{-}2$ which gives ${\cal M}_{n,n{-}2}(\{0,2\}) \sim z^{-n-2}$ which is indeed we observed in (\ref{MHVbar3}). For the $(n-3)$-line shift with ${\cal M}_{n,k}(\{a,b\})$, this overlap with $k=n{-}2$ for $\{1,2\}$ configuration and $k=n{-}3$ for $\{0,3\}$ configuration. Compared to the $(n{-}3)$ shift scaling table, it agrees as well. 

\subsection*{Risager shift}

The Risager multi-line shift \cite{Risager:2005vk} is a special case when we shift all $k$ negative helicity gravitons and leave all $n{-}k$ positive helicity gravitons unshifted, ie.
\begin{equation}
\widetilde{\lambda}_{\widehat{a}}(z) =  \widetilde{\lambda}_a + z\, \alpha_a\widetilde{\eta} \quad \mbox{for}\,\,\,\,a=n{-}k{+}1,\dots,n \label{shift}
\end{equation}
with the usual momentum conservation condition on $\alpha_a$. In our notation this is a $k$-line shift and $\{0,n{-}k\}$ helicity configuration. This is on the boundary of our phase space, but as for the other $\{0,n_1\}$ helicity configurations. An interesting observation is that the amplitude ${\cal M}_{n,k}$ evaluated on the special kinematics (\ref{prime6}) for $p=n{-}k$ vanishes
\begin{equation}
    {\cal M}_{n,k}(1',2',{\dots},n') = 0 \label{vanish}
\end{equation}
for the Risager shift. This is a very singular kinematical configuration when all $\widetilde{\lambda}_{a'}$ for positive helicity gravitons are collinear, and $[a'b']=0$ among any two of them. It can be shown using BCFW recursion relations that each term vanishes identically due to the vanishing helicity factors, and hence also the amplitude vanishes on this kinematics.

\subsection*{All $(0,n_1)$ configurations}

In our analysis of the $(n{-}2)$-line shift, we observed that the behavior for $z\rightarrow\infty$ for ${\cal M}_{n,k}(\{0,2\})$ was subleading with respect to other configurations, and the amplitude at infinity was equal to the same amplitude evaluated at special kinmatics (\ref{prime6}). It is natural to check the same statement for other $(n{-}n_1)$ shifts. Namely, we checked up to $n=9$ and $k=6$ that this statement holds for all non-trivial $(n{-}n_1)$,
\begin{equation}
        {\cal M}_{n,k} (\{0,n_1\}) = (-1)^{n+1} z^{n-2k+1} \times {\cal M}_{n,k} (1' ,2', \dots ,n') + {\cal O}(z^{n-2k})
    \end{equation}
The expression fails when $n_1=(n{-}k)$, and all positive helicity gravitons are unshifted. This is the Risager configuration. In that case, not only does the scaling decrease significantly, but the amplitude ${\cal M}_{n,k} (1' ,2', \dots ,n')=0$, as discussed above. 

\subsection*{More factorizations at infinity?}

A very natural question is if the factorization pattern extends to other $(n{-}n_1)$-line shifts. We can follow the same path as in the $n_1=2$ case and start with the $\overline{\rm MHV}$ analysis. If we choose $(a,b,c)=(1,3,4)$ and $(d,e,f)=(2,3,4)$ in the parity conjugate Hodges formula (\ref{Hodges1}), the leading term at infinity can be written as
\begin{equation}
    {\cal M}_{\rm inf} = \frac{\la12\ra\cdot H^8}{[12][13][1\widetilde{\eta}][23][2\widetilde{\eta}][3\widetilde{\eta}]^2}\times \frac{1}{\alpha_4^2}{\rm det}(\Psi_{134}^{234}) \label{higher}
\end{equation}
where in the $z\rightarrow\infty$ we can rewrite the determinant (and the $1/\alpha_4^2$ prefactor) to depend only on $(n{-}2)$ labels $(\omega,4,5,{\dots},n)$. At the same, the first term in (\ref{higher}) also is a particular representation of the four-point MHV amplitude for legs $(1,2,3,\eta)$ so it looks like ${\cal M}_{\rm inf} \sim {\cal M}_{4,2}(1,2,3,\eta) \times {\cal M}_{n{-}2,n{-}4}(\omega,4,5,{\dots},n)$, but the details are not correct. For the ${\cal M}_{4,2}$ piece is not permutational invariant because that would require a momentum conservation of four on-shell momenta, and $(p_1+p_2+p_3)^2\neq0$ so $p_1+p_2+p_3$ is not equal to some putative on-shell momentum $p_\eta$. Also, the form of the reduced Hodges matrix for labels $(\omega,4,5,{\dots},n)$ is not quite right. One possibility is to consider some deformed momenta $p_1$, $p_2$, $p_3$ which would satisfy on-shell momentum conservation, but that goes against the logic of having exactly these three labels undeformed. 

Hence, a more natural conjecture is to expand ${\cal M}_{\rm inf}$ as a sum of terms of the form ${\cal M}_{3,1}\times{\cal M}_{n{-}1,k{-}1}$. This form does not require the momenta $p_1$, $p_2$, $p_3$ to satisfy the on-shell momentum conservation. In fact, we observed for NMHV amplitudes $n=6,7,8$ that the following formula for the amplitude at infinity holds 
 \begin{equation}
        {\cal M}_{n,3}(\{1,2\}) = (-1)^n z^{n-4} \Bigg\{  s_{12} \times \frac{[2\tilde{\eta}]^8}{[12]^2 [1 \tilde{\eta}]^2 [2 \tilde{\eta}]^2} \times{\cal M}_{n-1,2}(\omega^+ 3'^- 4'^- 5'^+ \cdots n'^+) + (2 \leftrightarrow 3)\Bigg\} \label{higher2}
    \end{equation}
where $\omega$ is different in each term. This formula does not seem to generalize to other cases, but it might be just an accident or a tip of the iceberg of other more general expressions which would work for other $n,k$. We will leave this investigation for future work. 

\subsection*{Yang-Mills amplitudes at infinity}

We can compare the results for graviton amplitudes with the analogous behavior at infinity in the Yang-Mills theory. The $z$-scalings are different, and the analysis of MHV and $\overline{\rm MHV}$ formulas is particularly easy. But at higher $n,k$ we do find interesting results which are very similar to what we observed for graviton amplitudes. Namely, for the $(n{-}2)$-line shift we get, up to possible $(n,k)$-dependent overall signs,
\begin{align}
        {\cal A}_{n,k}(\{2,0\}) &= z^{3-k}\times {\cal A}_{\rm inf}(\{2,0\}) + {\cal O}(z^{2-k}) \label{res1}\\
        {\cal A}_{n,k}(\{1,1\}) &= z^{3-k}\times {\cal A}_{\rm inf}(\{1,1\}) + {\cal O}(z^{2-k}) \label{res2}\\
        {\cal A}_{n,k}(\{0,2\}) &= z^{2-k}\times {\cal A}_{n,k}(1',2',{\dots},n') + {\cal O}(z^{1-k}) \label{YM1}
\end{align}
where ${\cal M}_{n,k}(1',2',{\dots},n')$ is evaluated on the special kinmatics (\ref{prime6}) and the first two configurations exhibit the factorization pattern,
\begin{align}
 {\cal A}_{\rm inf}(\{2,0\})  &= \frac{1}{[12][1\widetilde{\eta}][2\widetilde{\eta}]}\times  {\cal A}_{n{-}1,k{-}1}(\omega^-,3'^-,{\dots},k'^-,(k{+}1)'^+,{\dots},n'^+)\nonumber\\
 {\cal A}_{\rm inf}(\{1,1\})  &= \frac{[2\widetilde{\eta}]^4}{[12][1\widetilde{\eta}][2\widetilde{\eta}]}\times  {\cal A}_{n{-}1,k{-}1}(\omega^+,3'^-,{\dots},(k{+}1)'^-,(k{+}2)'^+,{\dots},n'^+) \label{YM2}
\end{align}
which is very similar to (\ref{res3}) and (\ref{Infgen}). Also, the observation about (\ref{vanish}) for Risager shift does apply to Yang-Mills amplitudes too. This suggests that the observed behavior at infinity holds to both gluon and graviton amplitudes (up to different scalings which are important for the on-shell constructibility), and it should be possible to derive it from more general principles. 

\section{Conclusions and Outlook}

In this paper, we studied behavior of tree-level graviton amplitudes at infinity. Motivated by the multi-unitarity cut in the context of four-point gravity loop amplitudes, we mostly focused on the $(n{-}2)$-line anti-holomorphic shift. For MHV and $\overline{\rm MHV}$ amplitudes, we were able to derive precise formulas for the leading behavior at infinity. While the MHV amplitudes produce the same amplitudes evaluated at different kinematics, the $\overline{\rm MHV}$ amplitudes demonstrate an interesting form of factorization into three-point and $(n{-}1)$-point amplitudes. Both patterns generalize to the $n$-point N$^k$MHV case: two out of three helicity configurations factorize, while the third helicity configuration evaluates to the same amplitude on shifted kinematics. In fact, the third configuration has a vanishing leading term at infinity due to the helicity mismatch on the factorization, and hence the dominating hence (which is suppressed by one power of $z$) is the amplitude evaluated on special kinematics. We also discussed generalization for other $(n{-}n_1)$-line shifts but the factorization behavior is less clear there, and we will investigate it in the future.

Both patterns should naturally unify in the context of ${\cal N}{=}8$ SUGRA amplitudes, where the multi-line shift also involves the shift of the Grassmann variables (from the supermultiplet). We believe this could provide additional insights as it happened with the super-BCFW recursion relations \cite{Mason:2009sa,Arkani-Hamed:2009hub} which naturally evolved into the Grassmannian formulation for the ${\cal N}{=}4$ SYM amplitudes \cite{Arkani-Hamed:2009ljj,Mason:2009qx,Arkani-Hamed:2009nll,Arkani-Hamed:2016byb}.

A very interesting direction is also to study the generalized Risager deformation considered in \cite{Cachazo:2024mdn}, where different shift spinors $\widetilde{\eta}_i$ were considered. This led to very interesting formulas which exhibit different types of factorization properties. This probes a larger phase space of all possible shifts and the behavior of amplitudes at infinity in a new direction. 

Finally, we also plan to return back to our original motivation: study the UV sector of gravity loop amplitudes from the behavior on cuts. It was shown in \cite{Edison:2019ovj} that just the scaling of tree-level amplitudes at infinity (under the same shift) has surprising consequences for the scaling of the cut loop amplitude. Our work should provide a further input: not only the scaling at infinity, but also the explicit form of the amplitude at the leading order. Hence, revisiting the gravity loop problem is a natural project to tackle. 

\section*{Acknowledgments}

We thank Freddy Cachazo for very useful correspondence on the related topics. This work was supported by the DOE grant No.SC0009999 and the funds of the University of California.

\bibliographystyle{JHEP}
\bibliography{refs}

\end{document}